\documentclass[manuscript]{acmart}
\usepackage[utf8]{inputenc}
\usepackage{amsmath}
\usepackage{float}
\usepackage{algorithm}
\usepackage{algpseudocode}
\usepackage{caption}
\usepackage{enumitem}
\newtheorem{theorem}{Theorem}

\usepackage[utf8]{inputenc}
\usepackage{pgfplots}
\DeclareUnicodeCharacter{2212}{−}
\usepgfplotslibrary{groupplots,dateplot}
\usetikzlibrary{patterns,shapes.arrows}
\pgfplotsset{compat=newest}

\title{Partitioning and Deployment of Deep Neural Networks on Edge Clusters}
\author{Arjun Parthasarathy}
\author{Bhaskar Krishnamachari}
\date{September 2022}

\begin{document}

\begin{abstract}
    Edge inference has become more widespread, as its diverse applications range from retail to wearable technology. Clusters of networked resource-constrained edge devices are becoming common, yet no system exists to split a DNN across these clusters while maximizing the inference throughput of the system. Additionally, no production-ready orchestration system exists for deploying said models over such edge networks which adopts the robustness and scalability of the cloud. We present an algorithm which partitions DNNs and distributes them across a set of edge devices with the goal of minimizing the bottleneck latency and therefore maximizing inference throughput. The system scales well to systems of different node memory capacities and numbers of nodes, while being node fault-tolerant. We find that we can reduce the bottleneck latency by 10x over a random algorithm and 35\% over a greedy joint partitioning-placement algorithm, although the joint-partitioning algorithm outperforms our algorithm in most practical use-cases. Furthermore we find empirically that for the set of representative models we tested, the algorithm produces results within 9.2\% of the optimal bottleneck latency. We then developed a standalone cluster network emulator on which we tested configurations of up to 20 nodes and found a steady increase in throughput and decrease in end-to-end latency as the cluster size scales. In these tests, we observed that our system has multi-node fault-tolerance as well as network and system IO fault-tolerance. We have implemented our framework in open-source software that is publicly available to the research community at \url{https://github.com/ANRGUSC/SEIFER}.
\end{abstract}
\maketitle

\section{Introduction}
Deep Neural Networks (DNNs) have greatly accelerated machine learning across different disciplines, such as Computer Vision \cite{chai2021deep} and Natural Language Processing \cite{min2021recent}. Edge Inference is becoming an increasingly popular field with multiple facets \cite{wu2021accelerating}, as sensor-driven computation necessitates DNN inference in the field. Applications for Edge Inference range from retail to wearable technology \cite{chen2019deep, biswas2021survey}.

The edge can come in multiple configurations \cite{luo2021resource, sonkoly2021survey}, and there are multiple approaches to facilitate edge inference. For cloud-edge hybrid inference, one such approach is model compression \cite{gholami2021survey}, which deals exclusively with DNN optimization but does not address the system's runtime configuration. In this paper, we focus on clusters of resource-constrained edge devices. These \textit{edge clusters} are becoming increasingly common due to their low-cost and scalability at the edge \cite{premkumar2021survey}. Many lessons in high-availability and application portability can be taken from cloud computing \cite{sadeeq2021iot}. Unlike a cloud data center, the edge brings system resource limitations and communication bottlenecks between devices.

With this in mind, we address the following problem: \textbf{How can we take advantage of multi-device edge clusters to enable high-performance DNN inference while respecting computational resource constraints and taking into account the heterogeneity of communication links? Additionally, can we integrate cloud computing principles such as fault-tolerance, high-availability, and container-based abstractions to make edge inference production-viable?}

\begin{figure}
\centerline{\includegraphics[scale=0.6]{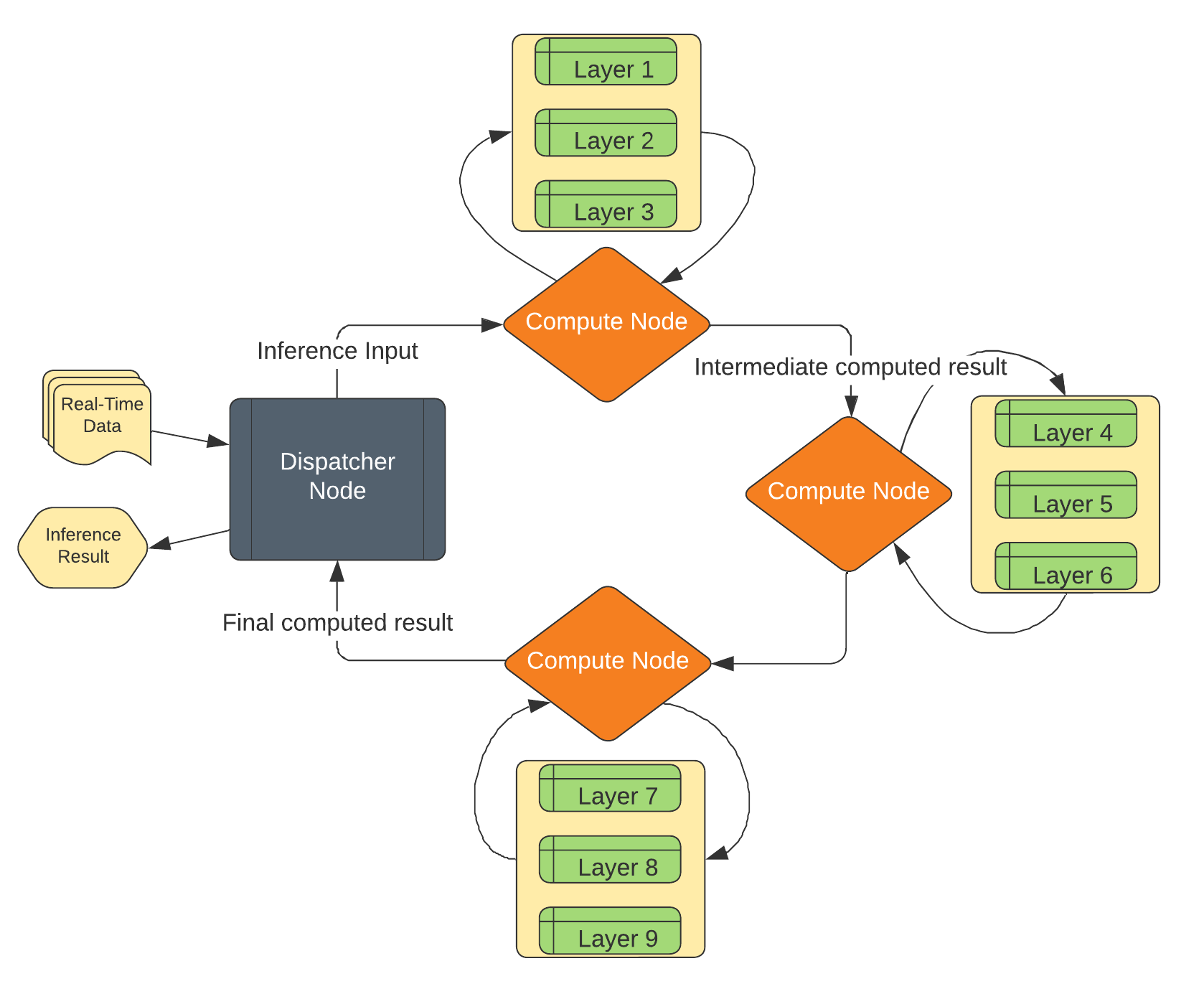}}
\caption{Partitioning and Distributing a Model Across Edge Devices to Create an Inference Pipeline}
\label{fig:partitioning}
\end{figure}

To partition a deep learning model, we first split the model into components that are executed sequentially. Each partition is assigned to a different edge device, and once each node performs inference with its piece of the model, that intermediate inference result is sent to the next node with the corresponding partition in the sequence. This inference pipeline is shown in Figure \ref{fig:partitioning}.

In an edge cluster, although we have a lower computational power in each node, we can take advantage of this inference pipelining to increase system throughput. Since each node can perform inference with its partition individually, prior nodes in the pipeline can send their finished inference results to the subsequent nodes in the pipeline and accept new batches. 

We define the \textit{throughput} metric of a system as the number of inference cycles it can perform per unit time. As we showed in our previous work DEFER \cite{parthasarathy2022defer}, we can achieve higher throughput with distributed edge inference as opposed to inference on a single device, providing that the node has enough capacity for the model.

The throughput is defined as the reciprocal of the \textit{bottleneck latency}. For nodes $[k] = \{1, 2, \dots, k\}$, the bottleneck latency $\beta$ is defined as

\begin{equation}
    \begin{array}{c}
        S = \{k \in [k] \mid c_k, \gamma_k\} \\
        \beta = \max_{s \in S}{s} 
    \end{array}
\end{equation}

where $c_k$ is the compute time of the operations on node $k$, and $\gamma_k$ is the communication time between node $k-1$ and $k$.

We use ResNet50 \cite{resnet50}, which is a representative model for our use case. On a Raspberry Pi 4, the inference speed was found to be 225 ms \cite{resnetinference}. Next, we found the amount of data transferred between each layer of the model. On average, 10.2 Mbits of data was transferred between layers. Given an average WiFi bandwidth of 6 Mbps for a low-end edge network, this gives us a communication time of $1.7s$. This is 7.5x slower than the compute time. In reality, many models are larger than ResNet50 and will therefore be split across devices, so each device will have less operations to execute. This means that communication time will outweigh compute time as the bottleneck. Therefore, we can simplify our bottleneck latency using Equation \ref{eq:bottleneck}.

\begin{equation}\label{eq:bottleneck}
    \beta = \max_{k \in [k]}{\gamma_k}
\end{equation}

Since throughput is defined as $\frac{1}{\beta}$, by minimizing the bottleneck latency we maximize inference throughput.

Additionally, we assume that all nodes are homogeneous in RAM. We discuss the different capabilities of edge devices in Section \ref{section:partitionmem}. If the devices are not the same capacity, then the algorithm will take the smallest memory capacity across all nodes in the cluster, and take that as the capacity of each node.

In this paper, we primarily analyze image and text models due to their prevalence on the edge for visual analytics applications \cite{8781894, nayak2021review}. We build on prior solutions by addressing both system resource limitations and cloud-computing features to create a robust inference framework.

We make two contributions:
\begin{enumerate}
    \item A partitioning and placement algorithm for DNNs across a cluster of edge devices distributed spatially within the same WiFi network. The algorithm finds the candidate partition points, finds the optimal partition sizes to transfer the least amount of data, and finds the arrangement of nodes with the highest bandwidth. Together, these aim to minimize the resulting bottleneck latency according to the throughput metric.
    \item A robust, containerized system to perform inference with the model partitions. The system is node fault-tolerant and dynamically updates the model partitions based on revisions to the model. The framework takes into account system resource limitations to provide a lightweight inference runtime. Our code is available at \url{https://github.com/ANRGUSC/SEIFER}.
\end{enumerate}

We found that our algorithm results in a 10x improvement over a random partitioning/placement algorithm, and a 35\% reduction in bottleneck latency for systems with 50 compute nodes. We  empirically observe an average approximation ratio of 1.092 for the bottleneck latency (i.e. it is 9.2\% more than the optimal bottleneck latency, on average). 

Additionally, we found that our containerized system has multi-node fault-tolerance and is able to recover from both network and system IO faults.

\section{Related Work}
\subsection{Edge Inference}
\subsubsection{DNN Model Slicing}

Some works mathematically perform DNN Model Slicing by layer \cite{zhang2020dynamic, zhang2022teeslice}, after calculating layer impact during the training stage. These do not account for communication demands on the edge. Others abstract model layers into certain "execution units," \cite{li2021slicing, cho2021dnn} which they then choose to slice based on certain resource requirements. Li \emph{et al.} \cite{li2018edge} regressively predict a layer's latency demand and optimizes communication bandwidth accordingly. These works are optimized for a hybrid edge-cloud pipeline and do not address the demands of a cluster of edge devices. Couper \cite{hsu2019couper} additionally evaluates model slices on a containerized platform and deploys these containers on an existing edge framework. However, it only addresses the case of a few sensors and edge devices, instead of a large edge cluster. This means that unlike our work, it does not have to optimize the placement of partitions onto devices and instead focuses on minimizing data transfer.

Our prior work DEFER addresses the partitioning and execution of DNNs on edge clusters, but does not address how to find candidate partition points or attempt to minimize bottleneck latency. Additionally, it does not leverage containerization and therefore is not easily portable between edge cluster configurations. We build on our prior work by introducing both containerization and an algorithm which aims to find optimal model partitions and node placement to minimize bottleneck latency.

Our paper builds on these works by addressing the bandwidth limitation of an edge cluster, and aims to maximize inter-node bandwidth to minimize bottleneck latency.

\subsubsection{Edge Inference Runtimes}
Intermittent edge inference \cite{gobieski2019intelligence} describes how to optimize edge inference for energy use on edge devices, but focuses on compression and pruning of model layers with a specialized inference runtime. Jupiter \cite{10.1145/3492323.3495630} orchestrates execution of a task on geographically distributed compute nodes based on a given task graph. This framework takes compute time as the bottleneck and uses a dynamic-programming solution to minimize computation time when distributing the task graph. Unlike this work, we take communication time as the bottleneck, which is more in line with real-world edge-cluster characteristics. Another edge task execution framework \cite{zhang2021joint} uses a multi-stage process to evaluate, schedule and containerize tasks on the edge. While creating an efficient inference runtime, this framework does not explicitly address the use case of DNN slicing across edge devices and therefore does not consider the bandwidth between nodes in the cluster. 

Our prior work DEFER is a standalone Python application that cannot scale across different node environments nor integrate the high-availability and fault-tolerance of a cluster framework. Couper, mentioned in the aforementioned section, creates a set of containers that can be run by another container orchestration framework. However, it doesn't construct the Kubernetes \cite{kubernetes} Pod execution units run on each node nor manage the lifecycle of the application components. Our framework pre-packages a Kubernetes distribution, allowing it to self-manage a standalone set of cluster resources and the scheduling of application components onto nodes. Broadly speaking, we introduce a containerized inference runtime optimized specifically for DNN partitioning and standalone cluster management based on cloud-computing principles, which prior edge inference runtimes do not address.

\section{Partitioning and Placement Algorithm}
We are given two graphs:

\begin{enumerate}
    \item An unweighted DAG $G_m$ representing the computation graph of a DNN, where each vertex represents a layer in the model. This DAG can be found using common ML libraries such as Tensorflow \cite{tensorflow} and Keras \cite{keras}.
    \item A weighted complete graph $G_c$ representing the communication graph of a cluster of homogeneous physical compute nodes, where each vertex represents a physical compute node and each edge represents the bandwidth between those nodes. The graph is complete because we assume that these edge devices will communicate over the same WiFi network.
\end{enumerate}

Our goal is to optimally partition the model and place these partitions on a set of edge devices. We do so as follows.

\subsection{Converting a Complex DAG to a Linear DAG}

\begin{figure}[htbp]
\centerline{\includegraphics[scale=0.4]{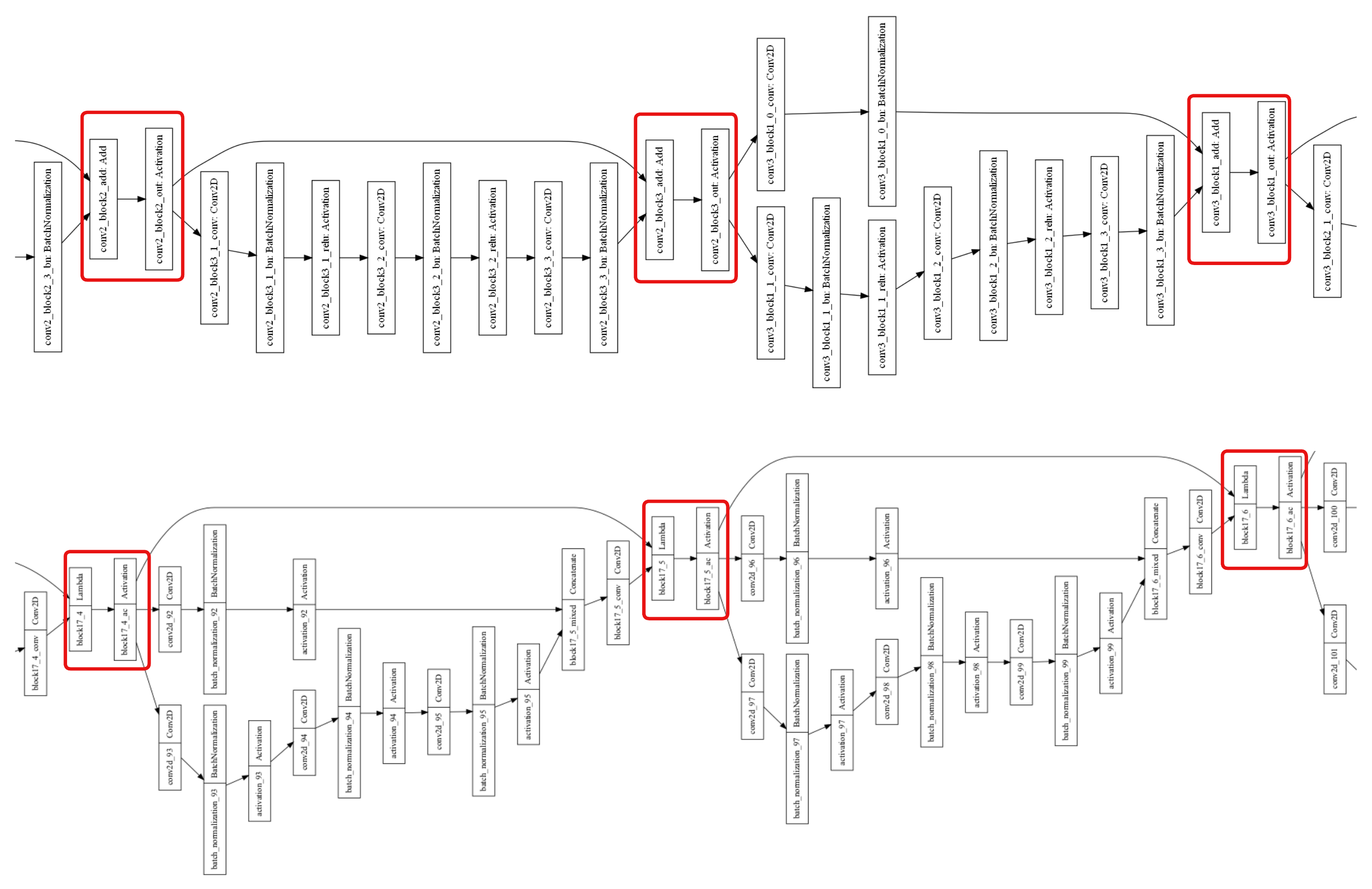}}
\caption{Partition points for ResNet50 and InceptionResNetV2 models}
\label{fig:examplepartpts}
\end{figure}

First, we need to distill $G_m$ into a linear DAG. The vertices where it is possible to partition the model are called ``candidate partition points." We illustrate this in Figure \ref{fig:examplepartpts}.

For $v \in V$, edges $e \in E$ and source vertex $s$ of $G_m$, find the longest path from $s$ to $v$. This can be done by topologically sorting the DAG and for each vertex in the resulting list, relaxing each neighbor of that vertex. We call the length of this longest path the \textit{topological depth} of that vertex in the graph. Let $LP(v)$ denote the length of longest path from $s$ to $v$.

To verify that all paths from vertex $v_{prev}$ go through vertex $v$, use a modified DFS by recursing on the incident edges of each vertex. If we encounter a vertex with a greater topological depth than $v$, return false. If we reach vertex $v$, return true. Let $AP(v_{prev}, v)$ denote the result of this algorithm.

Given the previously found candidate partition point $p_{k-1}$ and the current vertex $u$, the next candidate partition point $p_k = u$ iff:

\begin{enumerate}
    \item $LP(u) \neq LP(v) \forall v \in \{V - u\}$
    \item $AP(p_{k-1}, u) = \text{true}$
\end{enumerate}

with $p_0 = s$.

The time complexity of LP is $O(V+E)$. AP runs in polynomial time by returning upon reaching a vertex with a greater topological depth. Therefore, this algorithm runs in polynomial time.

Figure \ref{fig:examplepartpts} shows the candidate partition points at certain sections of the DAG of ResNet50 \cite{resnet50} and InceptionResNetV2 \cite{inceptionresnetv2}.

\begin{figure}[htbp]
\centerline{\includegraphics[scale=0.4]{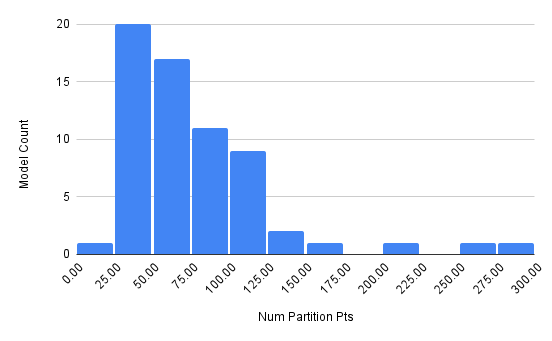}}
\caption{Histogram of number of candidate partition pts}
\label{fig:numpartpts}
\end{figure}

As shown in Figure \ref{fig:numpartpts}, almost all the models have at least 25 candidate partition points. This is more than enough granularity to split the model, given the upper bound described in section \ref{section:partitionmem} of the number of partitions we will need. There are some model architectures, like NASNet \cite{nasnet}, which do not allow partitioning under our scheme.

\begin{figure}[htbp]
\centerline{\includegraphics[scale=0.4]{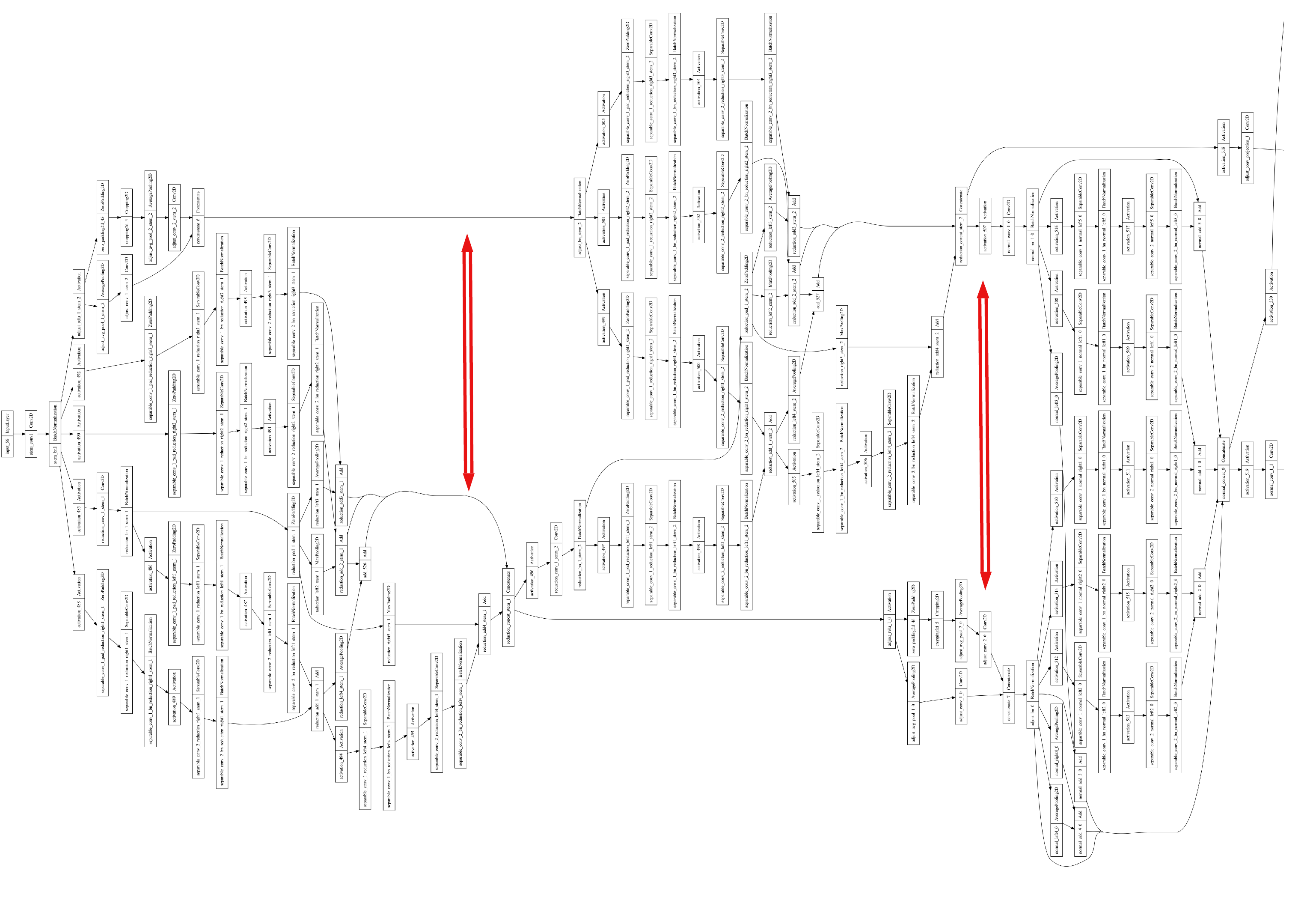}}
\caption{Portion of NASNet's layer DAG}
\label{fig:nasnet}
\end{figure}

As shown in Figure \ref{fig:nasnet}, NASNet cannot be partitioned because there is no single point that splits the model into a distinct execution unit that does not have any dependencies to a previous or subsequent layer. If we run our LP algorithm, we find that there is no single layer that has distinct topological depth from other layers. We found that 64 of the 66 (97 \%) pretrained Keras models could be partitioned under our scheme, and only the NASNet variants could not.

\subsection{Optimal model partitioning and placement}

Our goal is to maximize throughput of the system. As previously discussed, this means we need to minimize the bottleneck latency. Latency is defined as $\frac{\text{data}}{\text{bandwidth}}$. Given a tuple of partition points $P_{opt}$, their transfer sizes $T$, and a set of bandwidths $B$ between compute nodes, the latency between each set of compute node is defined as 

\begin{equation}\label{eq:latency}
    \gamma_k = \frac{T_{opt, k}}{B_k} \forall 0 \leq k < \lvert P_{opt} \rvert
\end{equation}

The bottleneck latency for the system is then given by Equation \ref{eq:bottleneck}. For the purposes of explanation, we separate the problems of optimizing the partitions (thereby optimizing transfer size) and optimizing placement (thereby optimizing bandwidth between nodes). We show empirically that this results in the the smallest bottleneck latency. In Section \ref{section:results}, we compare this formulation to an algorithm that tries to jointly optimize transfer size and bandwidth.

\subsubsection{Finding optimal partitions}
Our heuristic for finding optimal partitions is the ``transfer size" of the partition; i.e how much data will be transferred from that partition to the next. Given the tuple of candidate partition points $P = (p_0, p_1, \dots, p_k)$, we now need to find a set of model partitions which minimizes the sum of transfer sizes. Assuming a batch size of 1, the transfer size $t_k$ of candidate partition point $p_k$ is defined as

\begin{equation}
    t_k = \frac{\eta(p_k)}{\lambda}
\end{equation}

The function $\eta(p_k)$ finds the size of the output array of candidate partition point $p_k$.

$\lambda \approx 1.44 * 2.1$ represents the total compression ratio given by multiplying the average ZFP compression ratio \cite{zfp} by the average LZ4 compression ratio \cite{lz4}.

To better illustrate our algorithm, we classify the transfer size $t_k$ into 3 transfer size classes (``low'', ``medium'', or ``high'') based on the distribution of the transfer sizes. We discuss how many transfer size classes are actually necessary in section \ref{section:transfersizedist}.

\begin{equation}
\label{eq:classes}
\begin{array}{lrr}
C = \{L, M, H\} & t_k \in C & \forall 0 \leq k < |P|
\end{array}
\end{equation}

The optimal set of partitions is the scheme which minimizes the sum of the transfer sizes of said partitions.
\\
Let $G_p$ represent a DAG, where each vertex is represented by a possible partition. The vertices are defined as follows: 

\begin{equation}
\begin{array}{lr}
    V = \{\{p_i, p_{i+1}, \dots, p_j\} \mid \omega(\{p_i, p_{i+1}, \dots, p_j\}) < \kappa \} & \forall{0 \leq i < \lvert P_{opt} \rvert, 0 \leq j < \lvert P_{opt} \rvert - i}
\end{array}
\end{equation}

The set of vertices represents every possible contiguous subarray of candidate partition points, where $\omega(P)$ finds whether the memory use of partition $P$ is within the memory capacity $\kappa$ of the compute node. As discussed in Section \ref{section:partitionmem}, we quantize the models to reduce their memory footprint. However, when calculating the memory footprint of a partition, we do not consider this quantization. This means that we are conservative on partition size and in turn provide extra space on each device for the memory overhead from containerization. We provide a discussion on deriving the model's memory usage in section \ref{section:partitionmem}. Each partition is a set of layers that fall between the partition points $p_i$ and $p_j$.

The set of edges is defined as follows:

\begin{equation}
    E = \{(u, v) \mid (u, v) \in V, \rho(u_{\lvert u \rvert - 1}) = \rho(v_0) - 1 \}
\end{equation}

The function $\rho(\upsilon)$ finds the index of element $\upsilon$ in $P_{opt}$. There is an edge between vertices if the last partition point of $u$'s partition is adjacent in $P$ to the first partition point of $v$'s partition. For example, if $u = [1, 2]$, $v = [3, 4]$, and $P = (1, 2, 3, 4)$, then $(u, v)$ is an edge. Each edge has a weight $w(u, v)$ which corresponds to its transfer size class.

\begin{figure}[htbp]
\centerline{\includegraphics[scale=0.5]{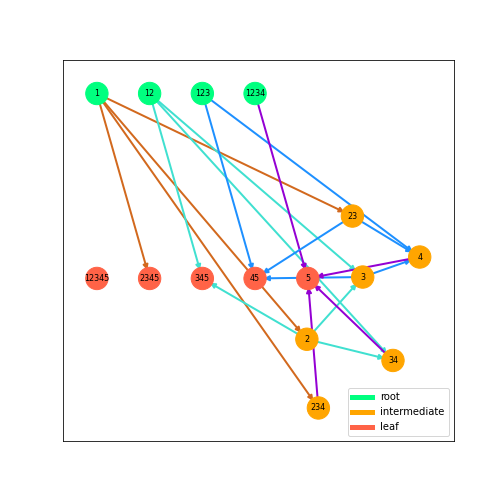}}
\caption{Example partition graph, where the partition points are $P = \{1, 2, 3, 4, 5\}$}
\label{fig:partitiongraph}
\end{figure}

Figure \ref{fig:partitiongraph} shows an example partition graph, where edges that are the same color will have the same weight. In the figure, ``root'' vertices have in-degree 0, ``leaf'' vertices have out-degree 0, and ``intermediate'' vertices have neither.

\begin{algorithm}
\caption{Optimal Partitioning}\label{alg:partitioning}
\begin{algorithmic}
\State{\textbf{//} Map to store memoized paths}
\State{$pathFrom \gets \Call{NEW-MAP}$} \label{line:memoize}
\Procedure{MIN-COST-PATH}{$G$, $v$}
\If {$v.children = \emptyset$}
    \State \Return{$v, 0$}
\EndIf
\State{$partitionLastLayer \gets v[v.length - 1]$}
\If{$partitionLastLayer \notin pathFrom$}
    \State{$paths \gets []$}
    \For {$c \in v.children$}  
        \State{$path, cost \gets \Call{MIN-COST-PATH}{G, c}$}
        \State{$paths \gets \Call{APPEND}{paths, (path, cost)}$}
    \EndFor
    \State{$pathFrom[partitionLastLayer] = \Call{MIN}{paths}$}
\EndIf
\\
\State{$minPath, minCost \gets pathFrom[partitionLastLayer]$}
\State{$chosenNode \gets minPath[0]$}
\State{\textbf{//} Path starting at v and going to a leaf}
\State{$newPath \gets \Call{APPEND}{[v], ...minPath}$}
\State{$newCost \gets minCost + w(v, chosenNode)$}
\State \Return{$newPath, newCost$}
\EndProcedure
\\
\Procedure{PARTITION}{G}
\State{$roots \gets \Call{GET-ROOT-VERTICES}{G}$}
\For {$r \in roots$}
    \State{$path, cost \gets \Call{MIN-COST-PATH}{G, r}$}
    \State{$paths \gets \Call{APPEND}{paths, (path, cost)}$}
\EndFor
\State{$minPath, minCost \gets \Call{MIN}{paths}$}
\State{\textbf{//} Prepend dispatcher partition $\Delta$ to beginning of optimal partitions array}
\State{$bestPath \gets \Call{APPEND}{[\Delta], ...minPath}$}
\State \Return minPath
\EndProcedure
\\
\State{$\Theta \gets \Call{PARTITION}{G_p}$}
\end{algorithmic}
\end{algorithm}

Algorithm \ref{alg:partitioning} finds the shortest path in the graph from a root to a leaf. Since edges which bridge the same candidate partition points (and have the same color as shown in Figure \ref{fig:partitiongraph}) will have the same subsequent paths, we can memoize the shortest path. On line \ref{line:memoize}, we store a map on which tells us for each candidate partition point what the shortest path is from that point. Using memoization, Algorithm \ref{alg:partitioning} takes $O(N)$ to find the shortest path, but $O(N^2)$ to construct the partition graph. Therefore the runtime of Algorithm \ref{alg:partitioning} is $O(N^2)$, where N is the number of nodes.

We also need to take into account the latency between the dispatcher node and the first compute node, since the model's input data is very large. After finding the array of optimal partitions, we prepend a special \textit{dispatcher partition} to the array of optimal partitions, which represents the runtime that sends inference input data. The transfer size corresponding to this partition is $\eta(p_0)$, where $p_0$ is the first candidate partition point and represents the model's input layer. We don't need to worry about the latency between the last compute node and the dispatcher node, because the size of a finished inference result is far smaller than an array of input data. We quantify this size difference in Section \ref{section:dispatchertransfersize}.

Let $\Theta$ represent the set of chosen partitions. For each subarray in $\Theta$, we take the last element of the subarray, add that to the list of partition points $Q$, and add its corresponding transfer size to the list $S$. The resulting list is then sorted based on the topological depth of each partition point, so that the partitions are executed in the order they appear in the model.

\subsubsection{Finding optimal model placement}
\label{section:modelplacement}
With the set of optimal partitions $Q$ and their corresponding transfer sizes $S$, we now need to ``match" them to the vertices of $G_c$. We know from Equation \ref{eq:classes} that every element of $S$ is a bandwidth class of $C$. Let $c(e)$ return the bandwidth class of a given edge of $G_c$. We use the following threshold function to classify each edge:

\begin{equation}
    \tau(X, t) = \left\{
    \begin{array}{lr}
        c(e) = C_{\arg X - 1}, & \text{if } e < t\\
        c(e) = X, & \text{if } e \geq t
    \end{array}
\right\}{\forall e \in E_c}
\end{equation}
If the edge is greater than or equal to the threshold, it will be classified as class $X$, otherwise it will be classified as the class in $C$ right below $X$. We provide a discussion on estimating the bandwidth distribution of $G_c$ in section \ref{section:avgbandwidth}. In order for our algorithm to work, we set the number of transfer size classes equal to the number of bandwidth classes.

\begin{figure}[htbp]
\centerline{\includegraphics[scale=0.5]{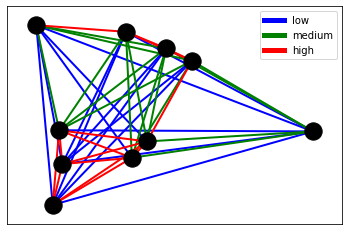}}
\caption{Example communication graph with different bandwidth classes}
\label{fig:commgraph}
\end{figure}

Figure \ref{fig:commgraph} shows an example communication graph. 

Given the array of transfer sizes $S$ and array of communication graph edges $E_c$, the lower bound on bottleneck latency we can achieve is given by Theorem \ref{thm:latencybound}.

\begin{theorem}\label{thm:latencybound}
    The lowest bottleneck latency we can achieve is:
    \begin{equation}\label{eq:minlatency}
        \min(\beta) = \frac{\max{S}}{\max{E_c}}
    \end{equation}
    Therefore, if we achieve $\min(\beta)$, then we have found the optimal minimum bottleneck latency.
\end{theorem}

We prove Theorem \ref{thm:latencybound} as follows:

Given the highest transfer size ($\max S$), then it must be matched with the highest bandwidth ($\max E_c$) to have the lowest bottleneck latency. There are two cases in which the system would have another bottleneck latency:

\begin{enumerate}
    \item 
        \begin{equation}\label{eq:highmismatch}
            \beta = \frac{\max{S}}{e}\forall e \in E_c - max(E_c)
        \end{equation}
    \item 
        \begin{equation}\label{eq:higherlatency}
            \beta = \frac{s}{e}\forall \{s \in S - max(S), e \in E_c - max(E_c) \mid \beta \geq \frac{\max{S}}{\max{E_c}}\}
        \end{equation}
\end{enumerate}

In Equation \ref{eq:highmismatch}, the latency of the system would be higher than Equation \ref{eq:minlatency}, since the transfer size is being matched with a lower bandwidth edge. In Equation \ref{eq:higherlatency}, some other transfer size $s$ and bandwidth $e$ may result in a higher bottleneck latency, in which case Equation \ref{eq:minlatency} still holds. Therefore, Theorem \ref{thm:latencybound} holds.

We run tests in Section \ref{section:results} to see how often we get this optimal solution. Algorithm \ref{alg:subgraphkpath} performs the matching between $S$ and $G_c$ to try to reach the optimal latency as outlined above. Let $N$ represent the array of nodes that we choose from $G_c$, with length $\lvert S \rvert$.

\begin{algorithm}
\caption{Finding K-Paths}\label{alg:subgraphkpath}
\begin{algorithmic}
\Procedure{SUBGRAPH-K-PATH}{$X$, $k$, $s$, $u$}
\State{\textbf{//} Sort by weight in descending order}
\State{$edgeList \gets \Call{SORT}{G_c, \{e \in E_c \mid w(e)\}, reverse=\text{TRUE}}$} \label{line:sortedgelist}
\State{$low \gets 0$}
\State{$high \gets edgeList.length$}
\State{$bestPath \gets []$}
\While{$low < high$}
    \State{$median \gets \frac{(low + high)}{2}$}
    \State{$\tau(X, edgeList[median])$}
    \State{$\tau(X, threshold)$}
    \State{\textbf{//} $G^X_c$ is the induced subgraph of $G_c$ with only bandwidth class $X$ edges}
    \State{$G^X_c \gets \{E^X_c = \{e \in E_c \land c(e) = X \mid e\} \mid V^X_c, E^X_c\}$}
    \State{$result \gets \Call{K-PATH}{G^X_c, k, s, u}$}
    \If{$result = \text{FALSE}$}
        \State{$low \gets median + 1$}
    \Else
        \State{$high \gets median$}
        \State{$bestPath \gets result$}
    \EndIf
\EndWhile
\For{$N \in bestPath$}
    \State{$\Call{DEL}{G_c, N}$}
\EndFor
\EndProcedure
\end{algorithmic}
\end{algorithm}

In algorithm \ref{alg:subgraphkpath}, we use the color-coding k-path algorithm \cite{alon1995color}, which finds a path of length $k$ (where $k$ is the number of vertices) in $G^X_c$ if a $k$-path exists and does so in polynomial time if $k < \log(|V^X|)$. See section \ref{section:partitionmem} for how we can bound the runtime of the algorithm. We use a binary search to find the maximum threshold for which a $k$-path exists. On line \ref{line:sortedgelist}, we sort in descending order so that we can find the maximum viable edge-weight threshold with a binary search. As $N$ starts to be filled in, the $k$-paths have to be found between certain nodes in order for $N$ to be a contiguous path of nodes. We modify the $k$-path algorithm to start at $s$ and stop once it reaches $u$. We make the algorithm more efficient by stopping a particular iteration if we reach $u$ before we have a path of length $k$. If $s$ is $null$, find any find any $k$-path that ends at $u$. Similarly, if $u$ is $null$, find any $k$-path that starts at $s$. 

\begin{algorithm}
\caption{K-Path Matching}\label{alg:kpathmatching}
\begin{algorithmic}[1]
\Procedure{K-PATH-MATCHING}{S, C}
\For{$X \in C$} \label{line:xmatch-begin}
    \State{$x\_paths \gets \Call{FIND-SUBARRAYS}{S, X}$} \label{line:findsubs}
    \State{$x\_paths \gets \Call{SORT}{x\_paths, \{p \in x\_paths \mid p.length\}}$}
    \For{$i \gets 0 \textbf{ to } x\_paths.length$}
       \State{$startIdx \gets \Call{INDEX-OF}{x\_paths[i][0], S}$} 
       \State{$startV \gets N[startIdx]$} \label{line:startV}
       \State{$endV \gets N[startIdx + x\_paths[i].length + 1]$} \label{line:endV}
       \State{$path \gets \Call{SUBGRAPH-K-PATH}{X, x\_paths[i].length + 1, startV, endV}$}
       \State{$N[startIdx:startIdx + path.length] \gets path$} \label{line:dots2}
    \EndFor
\EndFor \label{line:xmatch-end}
\EndProcedure
\end{algorithmic}
\end{algorithm}

\bigbreak

Algorithm \ref{alg:kpathmatching} performs the $k$-path matching of partitions onto vertices of $G_c$. 

FIND-SUBARRAYS() on line\ref{line:findsubs} returns a list of subarrays of a certain class, by iterating over the list of transfer sizes $S$.

Lines \ref{line:xmatch-begin}-\ref{line:xmatch-end} match paths for all bandwidth classes, starting with class $H$. On line \ref{line:startV}, $startV$ represents the vertex before the current $k$ path. If this is equal to $null$, meaning that the algorithm hasn't reached the iteration of finding the previous $k$-path, then SUBGRAPH-K-PATH will find a path that starts at any vertex. Similarly, on line \ref{line:endV} $endV$ represents the vertex of the current $k$-path. If this is equal to $null$, meaning that the algorithm hasn't reached the iteration of finding that subsequent $k$-path, then SUBGRAPH-K-PATH will find a path that ends at any vertex.

By starting with the longest $H$-subarrays and working to the shortest $L$ subarrays, we are greedily finding the best bandwidth paths to match with the highest transfer size terms of $S$. We continue this process until we have found $k$-path matchings for all subarrays of $S$. In some cases, a high number of bandwidth classes will prevent the algorithm from returning a result, because it has very few edges to choose from during each iteration of the matching. In this case, we can re-run the algorithm with fewer bandwidth classes.

\section{Cluster Architecture}

To run on a resource-constrained edge cluster, we propose modifications from our earlier implementation of the DEFER framework \cite{parthasarathy2022defer} which make the system more robust. Rather than using the full Kubernetes framework (which is not suitable for resource-constrained devices), we use microK8s \cite{microK8s}, which is adapted to run on the edge. MicroK8s only modifies the Kubernetes control plane and underlying infrastructure, so we still use Kubernetes constructs for our framework. We create Kubernetes services and take advantage of in-cluster DNS to allow pod communication independent of their lifecycle. In the event of node failure, pods can be rescheduled to healthy nodes and the system will continue running. 

\begin{figure}[htbp]
\centerline{\includegraphics[scale=0.1]{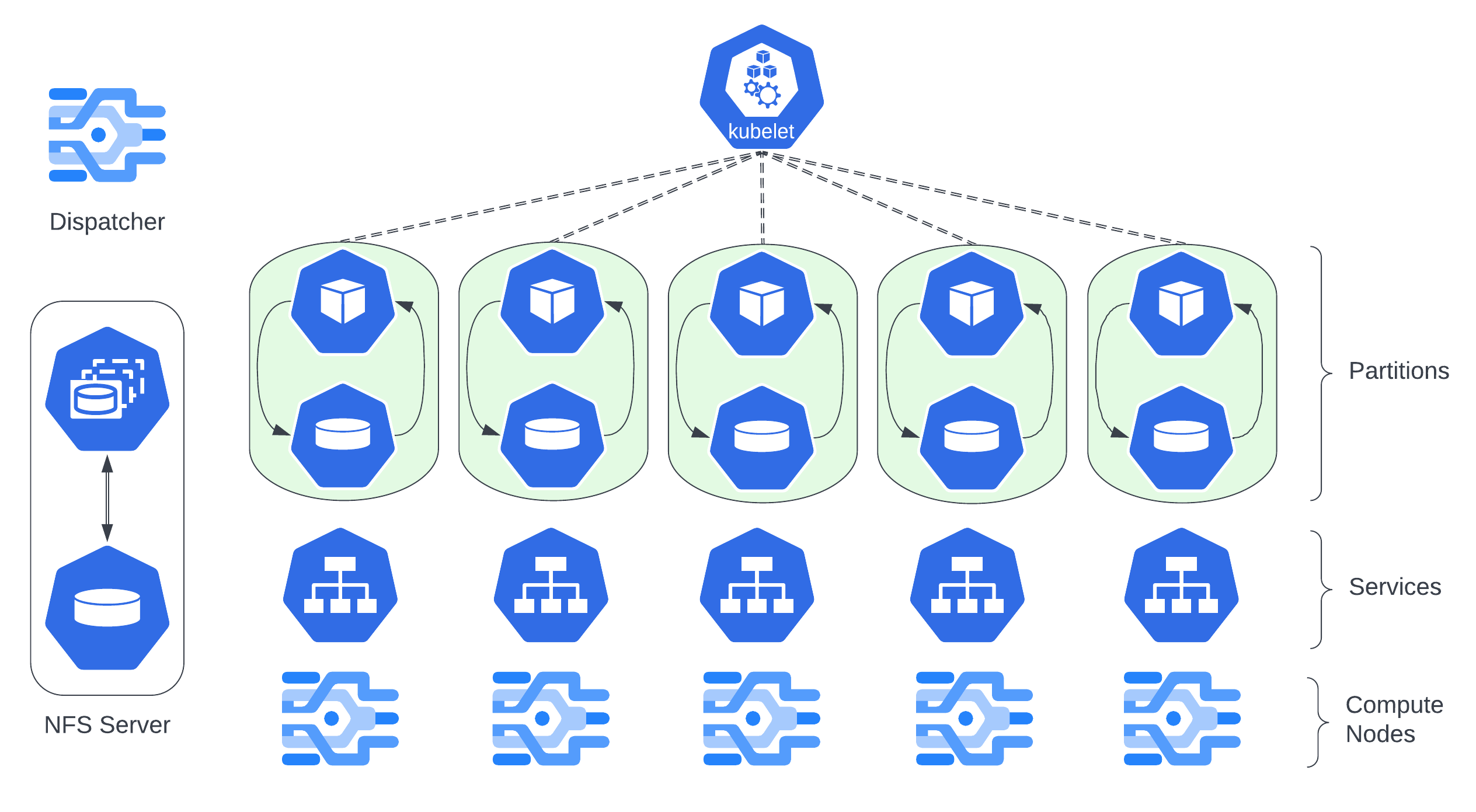}}
\caption{Kubernetes Cluster Overview}
\label{fig:systemoverview}
\end{figure}

As outlined in DEFER \cite{parthasarathy2022defer}, the system has the configuration step and inference step. We encourage readers to look at our prior work to understand how the configuration and inference steps work, because our objective here is to demonstrate the Kubernetes architecture of each step. We add an additional step, the \textit{system init step}, which configures the edge cluster and allows it to be run on any set of edge devices.

\subsection{System Initialization Step}
Upon system startup, the process of leader election starts and the \textit{Dispatcher Node} is chosen. The following events take place:

\begin{enumerate}
    \item \textbf{Scheduling IPerf Jobs}. The system initialization pod launches a job for each node which schedules a pod. Each pod contains a container which runs an IPerf \cite{iperf} server and an IPerf client, which it uses to find the bandwidth between itself and each other node in the cluster. Using a leader-follower architecture, the dispatcher directs each compute node when and where to connect in order to run the IPerf job. Each pod then directs the bandwidth info back to the dispatcher.
    \item \textbf{Scheduling Dispatcher Init Job}. A job is created for the pod that runs the partitioning and placement algorithm. It is scheduled onto the same node that the system initialization pod is running on, which has been chosen as the leader. The dispatcher pod is configured with the bandwidths between all nodes in the graph. 
    \item \textbf{NFS Server}. A cluster-wide NFS server is dynamically provisioned using NFS-Ganesha \cite{nfsganesha}. The NFS server will contain the files necessary to instantiate each node's partition. Since the NFS server has a lifecycle independent of each pod, it will preserve configuration data so that crashed pods can restart their inference runtimes.
\end{enumerate}

\begin{figure}[htbp]
\centerline{\includegraphics[scale=0.1]{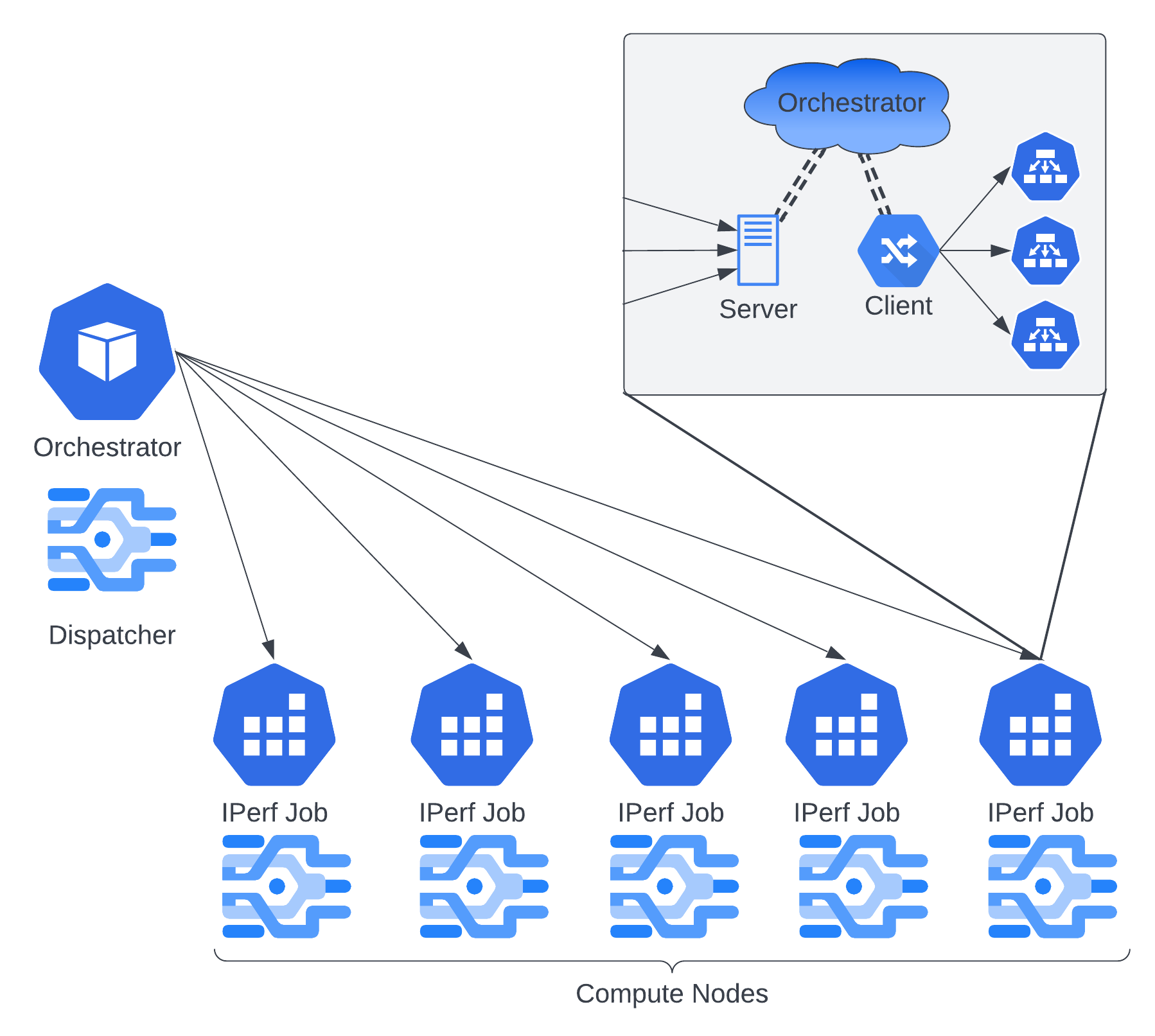}}
\caption{IPerf job orchestration to find bandwidths between nodes}
\label{fig:iperfjobs}
\end{figure}

\subsection{Configuration Step}
The pod within the dispatcher init job runs two containers which perform the DEFER configuration step as follows:

\begin{enumerate}
    \item \textbf{Partitioning Container}. The partitioning container pulls the model from a specified external repository. Using the stored node bandwidth data, the container runs the algorithm specified above to partition the model and assign each partition to a compute node. It quantizes the model (see Section \ref{section:partitionmem}) and saves the serialized model files to the NFS server.
    \item \textbf{Deploy Container}. The container creates a separate deployment for each inference pod to manage its lifecycle. Each inference pod is assigned to a certain compute node, and is configured to send its intermediate computed inference to another compute node. Additionally, it creates a deployment for the dispatcher to send and receive inference data during the inference step. This dispatcher deployment is scheduled to whichever node the placement algorithm scheduled the dispatcher partition.
\end{enumerate}

\subsection{Inference Step}

\subsubsection{Inference Pods}
Each inference pod has two containers:

\begin{enumerate}
    \item \textbf{Inference Runtime}. This container instantiates a TFLite \cite{tflite} model from the files on the NFS server. The container contains two FIFOs, which read and write serialized data to the IO container, respectively. Using ZFP \cite{zfp} and LZ4 \cite{lz4} compression, the runtime reads and decompresses data, runs it through the model, and then compresses and writes data. 
    \item \textbf{IO Container}. Contains two FIFOs and two TCP sockets. The FIFOs are used to read and write serialized data to the inference runtime, respectively. One TCP socket acts as a server and receives a connection from the previous compute node, while the other acts as a client and sends the computed inference data to the subsequent compute node. 
\end{enumerate}

\begin{figure}[htbp]
\centerline{\includegraphics[scale=0.1]{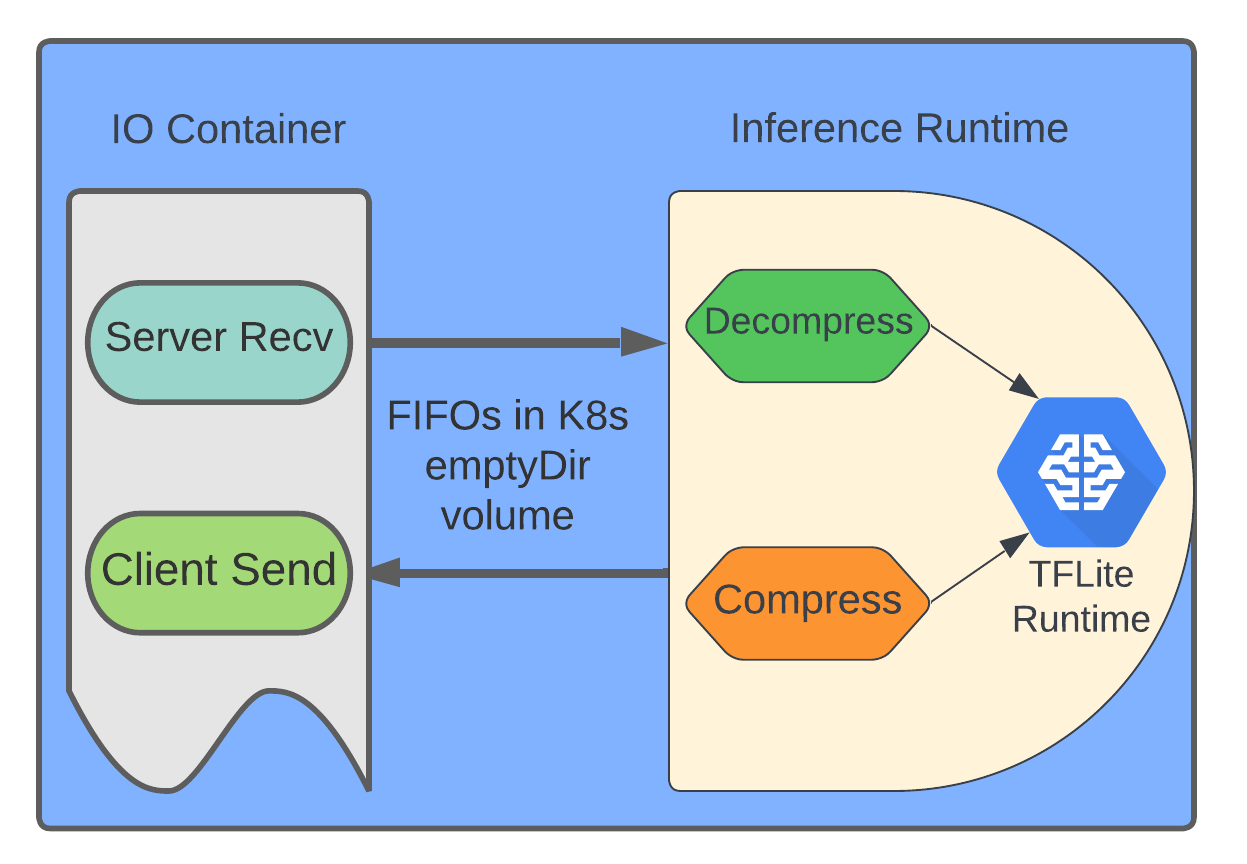}}
\caption{Inference Pod}
\label{fig:inferencepod}
\end{figure}

\subsubsection{Dispatching Inference Data}
Once the inference pods are deployed, the dispatcher runs three containers:

\begin{enumerate}
    \item \textbf{Processing Container}. Runs an HTTP server to read model input. It will convert the model input into the ZFP/LZ4 compressed form used in the system. The container contains two FIFOs, one to send model input data, and the other to receive finished inference results. The container runs an HTTP client which can be configured to send the finished inference data to a certain location.
    \item \textbf{IO Container}. Contains two FIFOs and two TCP sockets. The FIFOs are used to read and write serialized data to the processing container. One TCP socket acts as a server to receive finished inference results from the final compute node, and the other acts as a client to send model input to the first compute node.
    \item \textbf{Model Watch Container}. Watches for updates to the model on the external repository, and if it changes, the container will stop the inference pods and restart model partitioning. The cluster only needs to be shut down and restarted from the system initialization step if a new node is added.
\end{enumerate}

In lieu of running on microK8s, these containers can also be packaged and deployed on other edge inference frameworks \cite{akraino, azureiot, awsgreengrass}.

\begin{figure}[htbp]
\centerline{\includegraphics[scale=0.1]{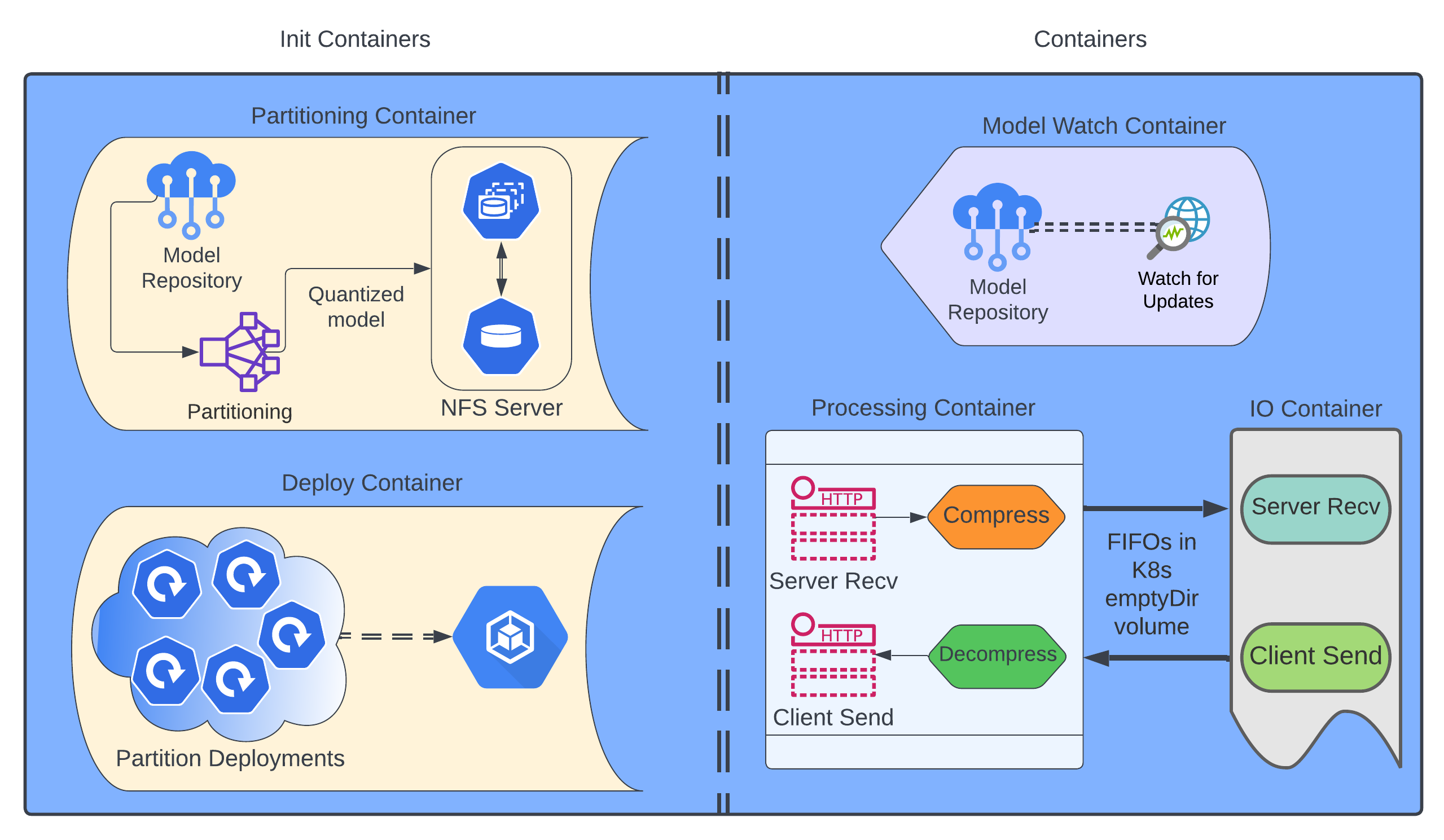}}
\caption{Dispatcher Pod}
\label{fig:dispatcherpod}
\end{figure}

\subsection{System Fault Tolerance}
Within each container, the IO mechanisms are all protected from failure. We elaborate on their recovery modes below.
\begin{enumerate}
    \item Network Failure
    \begin{enumerate}
        \item \textbf{IPerf job network failure} - system re-tries connection, until it reads a non-error state from the IPerf JSON output.
        \item \textbf{Client-side TCP connection error} - system checks for connection refused and DNS errors, and re-queries server until successful connection. If the desired pod is still in the ContainerCreating state, there will be a ConnectionRefusedError. If there was a node failure or the desired pod was restarted, there will be a DNSError while the service backing the desired pod waits for a new pod to be scheduled.
        \item \textbf{Connection Reset Error or EOF} - socket failed while reading data, so re-create the TCP server socket and wait for a new client connection.
        \item \textbf{Broken Pipe Error} - socket failed while writing data, so re-create the TCP client socket and wait for server to accept connection.
    \end{enumerate}
    \item File IO Failure
    \begin{enumerate}
        \item \textbf{Broken Pipe Error} - other end of FIFO failed while writing data, so pipe is re-created and opened for writing.
        \item \textbf{Connection Reset Error} - other end of FIFO failed while reading data, so pipe is re-created and opened for reading.
        \item \textbf{Blocking Error} - pipes are opened for writing in non-blocking mode, so if there's no incoming data on the pipe, the thread needs to wait for incoming IO.
    \end{enumerate}
    \item Kubernetes Pod Failure
    \begin{enumerate}
        \item \textbf{Pod Restart} - because of the RestartAlways PodFailure policy we selected, pods in the deployments will always restart after failure (ex. OOMKilled, Error, etc.).
    \end{enumerate}
    \item Node Failure
    \begin{enumerate}
        \item \textbf{Rescheduling Pods} - Pods will be evicted from their current node and rescheduled to a functioning node, where they will be started up again and configure themselves with TCP socket connections and NFS server data. For inference pods, the neighboring partitions will detect a break in the TCP socket connection and attempt to reconnect to the pod that was just rescheduled to a new node.
        \item \textbf{Rescheduling Volumes} - if the node containing the NFS server data goes down, then Kubernetes will need to dynamically provision a new PersistentVolume on another healthy node. However, this means that the current model partition files will be lost, and the cluster will need to be restarted. The in-built Kubernetes database, etcd, stores multiple copies of cluster configuration data across multiple nodes, allowing for cluster-wide configuration data to be node fault-tolerant. In our future work, we plan to explore sharding of our NFS server across multiple nodes to prevent single-node dependency. 
    \end{enumerate}
\end{enumerate}

\section{Measurements and Modeling}
In this section, we establish some baselines for the evaluation of our framework. In particular, we model the memory footprint of a model partition and some properties of Random Geometric Graphs which help us prove characteristics of the inter-node communication graph. The results from this section inform the configuration of our algorithm and our understanding of its efficacy.

\subsection{Memory Footprint of Partitions}
\label{section:partitionmem}

We used a sample of models from TFHub \cite{tfhub} for four different domains: image, video, audio, and text. Using TFLite \cite{tflite}, we performed float16 quantization and dynamic range quantization on each model. Float16 quantization has an insignificant accuracy drop and dynamic range quantization has a minimal accuracy drop, so neither of them should reasonably affect the accuracy of the model in a real-world scenario. 

\begin{figure}[htbp]
\includegraphics[scale=0.3]{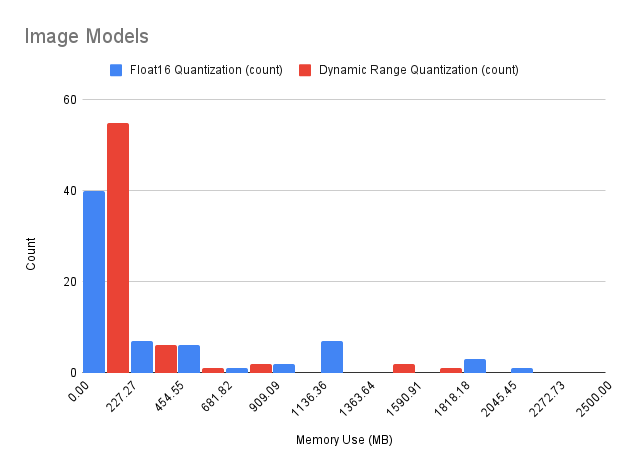}
\includegraphics[scale=0.3]{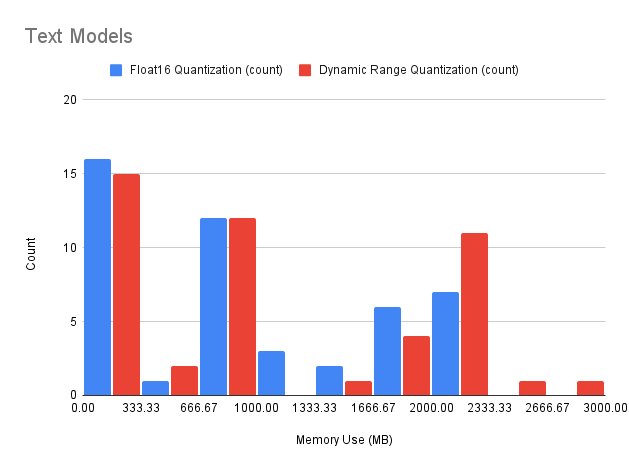}
\includegraphics[scale=0.3]{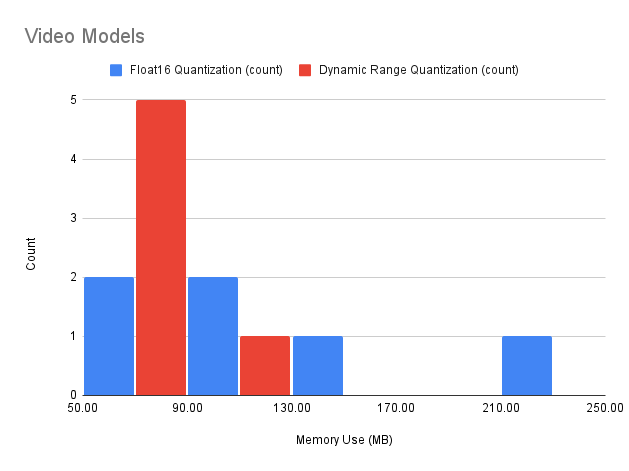}
\includegraphics[scale=0.3]{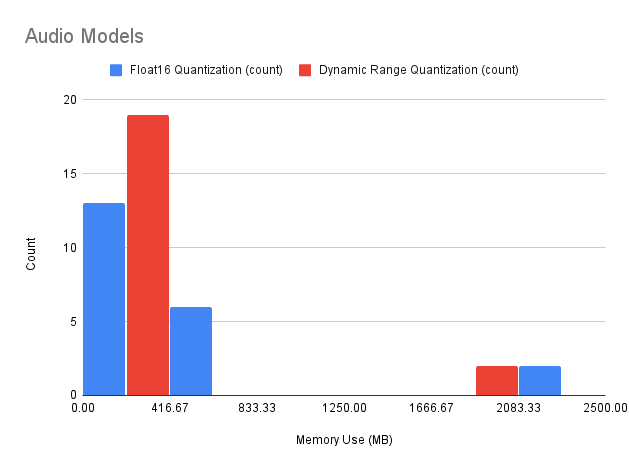}
\caption{Peak Memory Usage of Popular TFHub models}
\label{fig:memoryusage}
\end{figure}

We calculated the peak memory usage of each quantized model using the TFLite Model Benchmark Tool \cite{tflitebenchmark}, and the results are shown in Figure \ref{fig:memoryusage}. We focus on image and text models since they are the most memory-intensive and common use-cases for edge inference. By using float16 quantization, we can limit text models to 2000 MB peak memory usage. By using dynamic range quantization, we can limit image models to 2000 MB peak memory usage, and the vast majority of models will use less than 1000 MB. Table \ref{tab:numdevices} quantifies how many devices on average we would need to accommodate the model for each category of edge capability and memory capacity.

\begin{table*}[!htp]\centering
\caption{Number of Devices Necessary to Accommodate Models}\label{tab:numdevices}
\begin{tabular}{|c|c|c|c|}
\hline
&Low-End Edge (512 MB) &Mid-End Edge (1 GB) &High-End Edge (8 GB)\\
\hline
Image &3 &2 &1 \\
\hline
Text &\textbf{4} &3 &1 \\
\hline
\end{tabular}
\end{table*}

For our purposes, low-end edge is a Raspberry Pi Zero \cite{rpizero} with 512 MB of RAM, mid-end edge is the Raspberry Pi 3 Model B \cite{rpi3}, with 1 GB of RAM, and high-end edge is the Raspberry Pi 4 Model B \cite{rpi4} with up to 8 GB of RAM. We make this estimation assuming that extra memory will be used to run a MicroK8s control plane architecture. We can infer from our sampling that we will not need more than 4 low-end edge devices to accommodate a real-world model. In section \ref{section:modelplacement}, we said we could bound the runtime of the $k$-path algorithm. The runtime of the algorithm is roughly $O(4.32^k)$. Since we just found that there will be at most 4 partitions, we will only never need to find a path of length $k \leq 4$. Therefore, we can cap the runtime of the $k$-path algorithm at $O(4.32^4) < 350$ operations, which runs on the order of milliseconds.

\subsection{Distribution of Partition Transfer Sizes}
Using the same sample of Keras models, we partitioned the models and calculated the transfer sizes of all the candidate partition points.

\subsubsection{Transfer Sizes between Model Partitions}
\label{section:transfersizedist}

\begin{figure}[htbp]
\centerline{\includegraphics[scale=0.3]{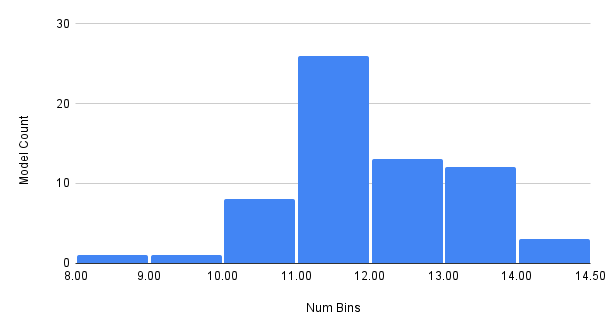}}
\caption{Histogram of number of bins necessary to represent each model's transfer sizes}
\label{fig:transfersizebins}
\end{figure}

Figure \ref{fig:transfersizebins} highlights the histogram of the number of bins, according to Doane's estimator, that each model's transfer sizes would need to adequately be represented in a histogram-like format. This number of bins is roughly equal to the number of transfer size classes necessary to represent the variation in each model's transfer sizes.

We can see that most of the models require 11 transfer size classes, and almost all the models requiring 11-13 classes. This informs our methodology of the number of transfer size classes that we choose in Section \ref{section:results}. In that section, we test an approximately equal number of transfer size classes below and above this number of 11, to see how it affects the bottleneck latency.

\subsubsection{Transfer Sizes between Dispatcher and Model Partitions}
\label{section:dispatchertransfersize}

The transfer size between the dispatcher and the first model partition is the input size of the model. The transfer size between the last partition and the dispatcher is the size of the computed inference result. For ResNet50 with a batch size of 1, the input is a 3-channel 224x224 pixel image. This comes out to an array size of 150328. The output is a soft-max array of 1000 image classes. This comes out to an array size of 1000. The input size is more than 100x the output size, and this factor is higher for many models that accept larger images but have a similar amount of output image classes. Therefore, the transfer size from the last partition to the dispatcher is negligible compared to the transfer size from the dispatcher to the first partition.

\subsection{Properties of Communication Graphs}
The wireless communication graph formed by a randomly deployed set of nodes can be modeled as an Erdős-Rényi Random Geometric Graph \cite{penrose2003random}, so we can extract certain characteristics from it.

\subsubsection{Average Bandwidth of an Edge Connection}
\label{section:avgbandwidth}

Let $B = 150$ be the range of the WiFi router in meters. We derive the equation for bandwidth given distance from the router based on Shannon's capacity equation ($C = \log_{2}{(1+\frac{S}{N})}$). Here, we assume that the signal decays proportionally to the inverse square of the distance between the device and the router. Equation \ref{eq:bandwidth} represents the bandwidth of a connection with a device $d$ meters from the router.  

\begin{equation}
\label{eq:bandwidth}
    D(x)=\log_{2}\left(1+\frac{a}{d^2}\right), d \in (1, B)
\end{equation}

In Equation \ref{eq:bandwidth}, we found $a=283230$ by assuming that the bandwidth at 80 m from the router was 5.5 Mbps, which matches the characteristics of a low-power edge network. Then, we can plug in the distance formula to derive the bandwidth given a position (x, y) on the 2D plane:

\begin{equation}
\begin{array}{lr}\label{eq:bwdistance}
r(x, y)=\log_{2}\left(1+\frac{a}{\sqrt{x^2+y^2}^2}\right)=\log_{2}\left(1+\frac{a}{x^2+y^2}\right) & x, y \in (-B, -1) \cup (1, B)
\end{array}
\end{equation}

In Equation \ref{eq:bwdistance}, we define our function on the domain $x, y \in (-B, -1) \cup (1, B)$. We do this for two reasons: to satisfy the domain of Equation \ref{eq:bandwidth}, and to simplify the creation of our geometric graphs for our simulations. From a practical standpoint, this means that we assume that no devices will be within 1 m of the router. Let two continuous random variables $X, Y \sim \text{Unif}(-B, B); X, Y \notin (-1, 1)$. Then their PDFs are

\begin{equation}
\begin{array}{lr}
f_X(x) = f_Y(y) = \frac{1}{2(B-1)} & x, y \in (-B, -1) \cup (1, B)
\end{array}
\end{equation}

Since $X$ and $Y$ are independent, their joint PDF is given by

\begin{equation}
\begin{array}{lr}
f_{XY}(x, y) = \left(\frac{1}{2\left(B-1\right)}\right)^{2} & x, y \in (-B, -1) \cup (1, B)
\end{array}
\end{equation}

Now, we can find the expected value of the transformation $r(x, y)$ over $X$ and $Y$.

\begin{equation}
\begin{array}{lr}
E[r(X, Y)] = \int_{-B}^B\int_{-B}^B r(x, y)f_{XY}(x, y)dxdy & x, y \notin (-1, 1)
\end{array}
\end{equation}

To find the standard deviation, we need to also calculate $E[r(X, Y)]^2$.

\begin{equation}
\begin{array}{lr}
E[r(X, Y)^2] = \int_{-B}^B\int_{-B}^B r(x, y)^{2}f_{XY}(x, y)dxdy & x, y \notin (-1, 1)
\end{array}
\end{equation}

We can then calculate the mean, standard deviation, and coefficient of variation.

\begin{equation}
\begin{array}{l}
\mu = E[r(X, Y)] \approx 4.766 \\
\sigma = \sqrt{E[r(X, Y)^2] - E[r(X, Y)]^2} \approx 1.398 \\
CV = \frac{\sigma}{\mu} \approx 0.293
\end{array}
\end{equation}

The average bandwidth between any two nodes in our communication graph is 4.766 Mbps and that the distribution of bandwidths in a randomly generated graph is relatively tight. We use this result in the next section to confirm the efficacy of our $k$-path algorithm.

\subsubsection{Clustering in RGGs}
\label{section:rggclusters}

We can model the induced subgraph of high edges $G^H_c$ with a random geometric graph.

Consider the case where we only have $L$ and $H$ bandwidth classes, and we want to split the graph such that all edges above the average bandwidth are classified as $H$.

Since the average bandwidth from Section \ref{section:avgbandwidth} was found to be $\mu \approx 4.766$, we find the distance from the router at which this bandwidth would be achieved using Equation \ref{eq:bandwidth}.

\begin{equation}
\begin{array}{l}
D(x) = \mu \\
x \approx 103.944
\end{array}
\end{equation}

Because we take the range of the WiFi router to be $B=150$, we can scale this result to the region $r \in [0, 1]^2$ necessary for an RGG.

\begin{equation}
r = \frac{103.944}{B} \approx 0.693
\end{equation}

Now, we can compute the proportion of vertices in the largest cluster and the \textit{cluster coefficient} of the graph \cite{dall2002random}. First, we find the average degree $\alpha$ of the graph given $r$. Let $N = \lvert V \rvert$ represent the number of vertices in $G_c$.

\begin{equation}
\begin{array}{l}
    a = \frac{\pi^{\frac{d}{2}}r^{d}}{\Gamma\left(\frac{d+2}{2}\right)} \\
    b = 2^{d}a \\
    \alpha = N b
\end{array}
\end{equation}

where $d = 2$ is the number of dimensions. We can find the proportion $P$ of vertices in the largest cluster (connected component) of the graph based on $\alpha$.

\begin{equation}
    P\left(\alpha\right)=1-\frac{1}{\alpha}\sum_{n=1}^{\lvert V^H \rvert}\frac{n^{\left(n-1\right)}}{n!}\left(\alpha e^{-\alpha}\right)^{n}
\end{equation}

Let's consider two practical cases, where $N = 10$ and $N = 50$.

\begin{equation}
\begin{array}{ll}
    N = 10 & N = 50\\
    \alpha \approx 60.343 & \alpha \approx 301.715\\
    P(\alpha) = 1 & P(\alpha) = 1
\end{array}
\end{equation}.

In both of these cases, $P(\alpha) = 1$ means that all of the vertices in the graph are part of the largest cluster, i.e they are all connected. This means that we are guaranteed to find a $k$-path of length $k \leq N$ in $G^H_c$.

We can also find the \textit{cluster coefficient} $C$ of the graph, which is a measure of how ``cliquish'' the graph is. This number is independent of $N$.

\begin{equation}
\begin{array}{l}
     H\left(x\right)=\frac{1}{\sqrt{\pi}}\sum_{i=x}^{\frac{d}{2}}\frac{\Gamma\left(i\right)}{\Gamma\left(i+\frac{1}{2}\right)}\left(\frac{3}{4}\right)^{\left(i+\frac{1}{2}\right)} \\
     C = 1 - H(1) \approx 0.587 \\
\end{array}
\end{equation}

This means that given any two vertices $i, j \in V$, if for some common vertex $k$, $(i, k) \in E$ and $(j, k) \in E$, then $(i, j) \in E$ with probability $\approx 0.587$. 

This indicates that the subgraph of bandwidth class $H$ edges exhibits cliquish behavior, and due to the high probability of cliquish edges, there is a large variety of $k$-paths in the subgraph.

\section{Evaluation Methodology}
\subsection{Algorithm Simulations}
\label{section:simulations}
We simulated a set of randomly placed edge devices using a random complete graph. For each evaluation, we created a random complete graph by drawing the positions of the nodes from a uniform distribution with the range $(-B, -1) \cup (1, B)$ used in Equation \ref{eq:bwdistance}. Between each set of nodes, we calculated the edge weight according to Equation \ref{eq:bwdistance}.

For each model, we ran Algorithm \ref{alg:kpathmatching} with a certain number of nodes, number of bandwidth classes, and node memory capacity. We used the set of nodes [5, 10, 15, 20, 50]. We used the set of bandwidth classes [2, 5, 8, 11, 14, 17, 20]. We used the set of node memory capacities [64, 128, 256, 512]. 

We used the following configuration to test the algorithm:

\begin{enumerate}
    \item \textbf{Number of Nodes} - 5, 10, 15, 20, or 50 randomly placed edge devices.
    \item \textbf{Number of Bandwidth Classes} - 2, 5, 8, 11, 14, 17, or 20 bandwidth classes, which provide granularity in how to classify the transfer sizes and edge bandwidths.
    \item \textbf{Node Memory Capacity} - 64, 128, 256, or 512 MB of RAM for a compute node.
\end{enumerate}

For each test, we used a different random communication graph generated using the procedure above. With each algorithm result, we then calculated the bottleneck latency according to Equation \ref{eq:latency}. The resulting bottleneck latency from each configuration of model, node capacity, number of nodes, and number of bandwidth classes was run 50 times and averaged. 

We compare the resulting bottleneck latency of our algorithm to that of the following two algorithms:

\begin{enumerate}
    \item \textbf{Random Algorithm} - Select a random node and a random partition that can be accommodated on that node.
    \item \textbf{Joint-Optimization Algorithm} - 
    Let $Q$ and $N$ represent the optimal set of partitions and optimal arrangement of nodes, respectively, chosen under this algorithm
    For each node $n$, do the following:
    \begin{enumerate}
        \item At each step choose the partition with the smallest transfer size that will fit within the node. Add this partition to the set of chosen partitions $p$.
        \item Starting at $n$, find the neighbor in the communication graph whose edge $e$ has the highest bandwidth, and add that to the path of chosen nodes $c$. Then, find the highest bandwidth edge from $e$, and so on.
        \item Compare the bottleneck latency found with $p$ and $c$ to the smallest bottleneck found with all nodes $n$ thus far, and update $Q$ and $N$ with $p$ and $c$ if the current bottleneck is smaller.
    \end{enumerate}
\end{enumerate}

For each of these algorithms, we used the same configuration and methodology as above to find the bottleneck latency. These algorithms don't use bandwidth classes, so we didn't need to include that as part of the configuration.

We also ran 1000 tests of the InceptionResNetV2 model with a random communication graph with 64 MB node memory capacity, 50 compute nodes, and 20 bandwidth classes. We found that the model reaches the optimal latency (as defined in Theorem \ref{thm:latencybound}) 54 times. This is a percentage of 5.4\% at optimality.

\subsection{Cluster Simulations}
We tested our system by creating a virtual cluster on a single host machine. We generate configurations of communication graphs commonly found in the real world.

\subsubsection{Graph Configurations}

\begin{enumerate}
    \item \textbf{Number of Nodes} - 5, 9, or 20 compute nodes
    \item \textbf{Node Arrangement} - Ring, Grid, or Cluster shape
\end{enumerate}

\begin{figure}[htbp]
\centerline{\includegraphics[scale=0.5]{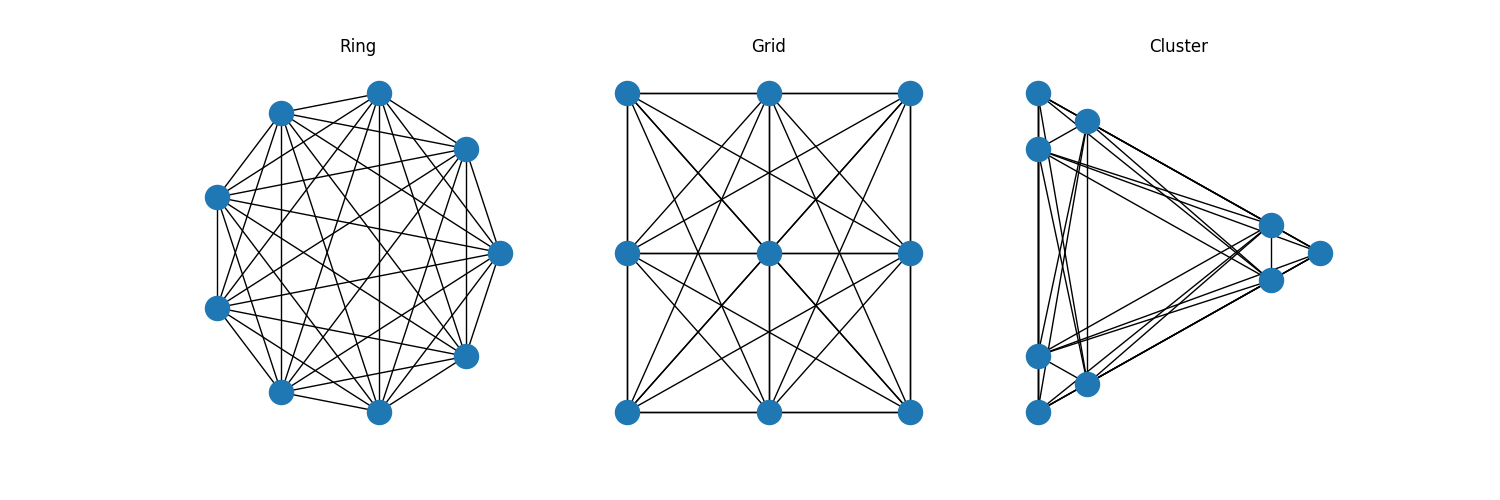}}
\caption{Communication graph configurations for a system with 9 compute nodes}
\label{fig:graphshapes}
\end{figure}

Figure \ref{fig:graphshapes} shows the different configurations that we test for a system of 9 compute nodes. In ChaosMesh, we use Equation \ref{eq:bwdistance} to define the bandwidth between a pair of nodes according to their distance apart in the graph.

\subsubsection{Test Environment Architecture}

To simulate an edge cluster on a single machine, we used Minikube \cite{minikube}. Since Minikube has a minimum node memory requirement of 1800 MB, we artifically restrict Algorithm \ref{alg:kpathmatching} to partition based on 64 MB node capacity so we can mimic an edge device. To simulate different network bandwidths between nodes, we used ChaosMesh \cite{chaosmesh}, which uses the TC-TBF \cite{tctbf} Linux algorithm. We've packaged our test environment as part of our code release.

\begin{figure}[htbp]
\centerline{\includegraphics[scale=0.5]{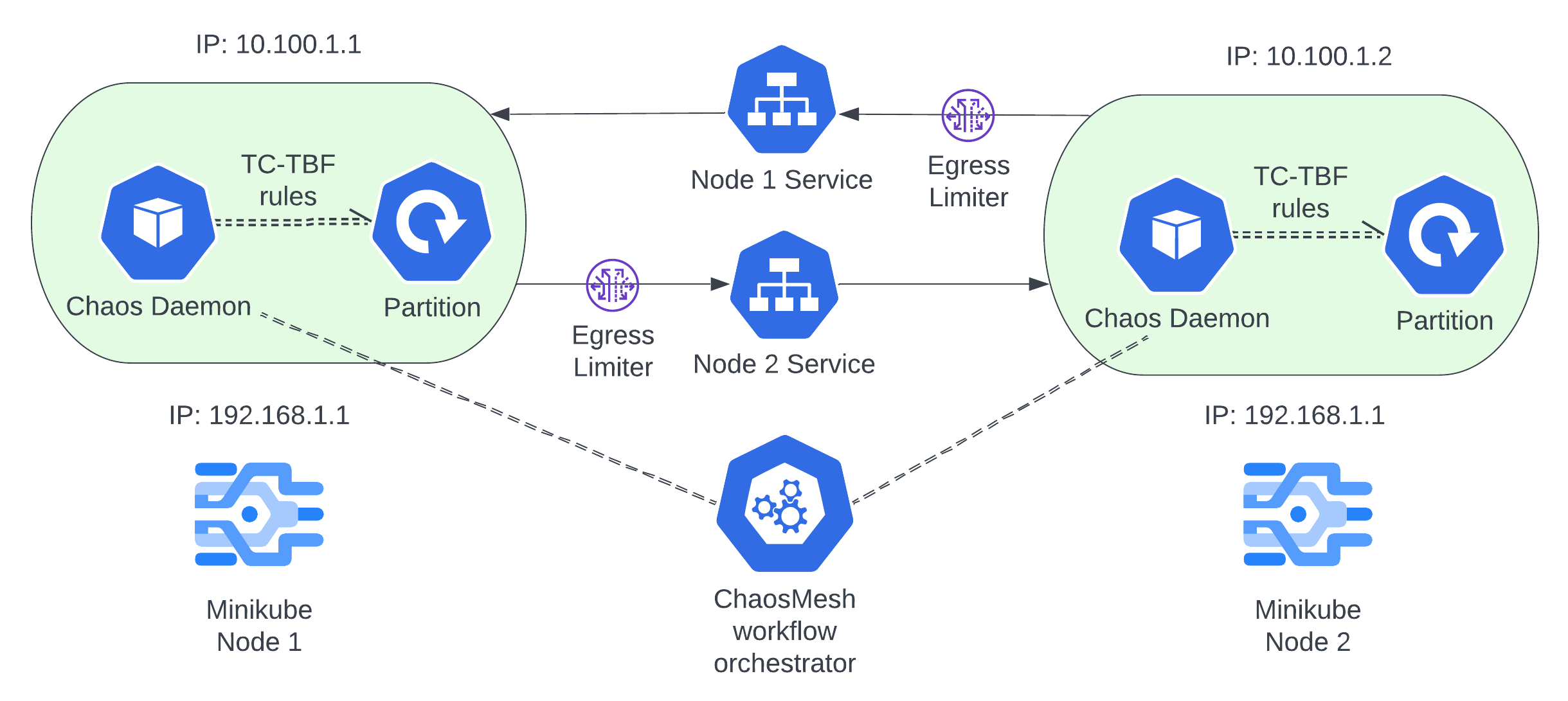}}
\caption{Test Environment Architecture}
\label{fig:testenv}
\end{figure}

Figure \ref{fig:testenv} depicts the architecture of the test environment. Each node gets a workflow of different \textit{NetworkChaos} rules. Each NetworkChaos rule specifies a TC-TBF bandwidth limit. Within TC-TBF, we control 3 parameters:

\begin{enumerate}
    \item \textbf{Rate} - the bandwidth in kbps, derived from our communication graph configurations in Figure \ref{fig:graphshapes}
    \item \textbf{Limit} - max number of TCP packets that can be queued on the sender side. It's recommended that this number be set to $2*\text{rate}*\text{latency}$. Given a max bandwidth of $18 mbps$ from Equation \ref{eq:bwdistance} and a conservative estimate of $2s$ for latency, the limit should be 10 MB.
    \item \textbf{Buffer} - max number of tokens that can be sent instantaneously. If this number is too high, it will allow high burst speeds during the IPerf bandwidth jobs shown in Fig. \ref{fig:iperfjobs} and result in an abnormally high bandwidth reading. If this number is too low, not enough tokens will be available at a time to send data, resulting in packet loss. In our testing, we found that a burst of 10 KB allows for accurate bandwidth readings with IPerf.
\end{enumerate}

Since TC-TBF only affects the egress bandwidth from each pod and uses the IP of the target pod to limit the bandwidth, we need to make sure that TCP packets exiting the pod have the correct source and destination IPs. Since a regular Kubernetes Service has its own cluster-wide IP and forwards packets to the pod it backs, the TC-TBF rules wouldn't take effect for any packets since they would have the destination IP of the \textit{service} which backs the next inference pod, rather than the IP of the pod itself. Therefore, our test environment uses Kubernetes \textit{headless services} to back each inference pod. Rather than returning the service's cluster-wide IP, a DNS query for a headless service will directly return the IP of the pod it backs, allowing TCP packets sent to another pod to have the correct destination IP.

\section{Results}
\label{section:results}
\subsection{Simulation Results}

\begin{figure}[htbp]
\centerline{\includegraphics[scale=0.5]{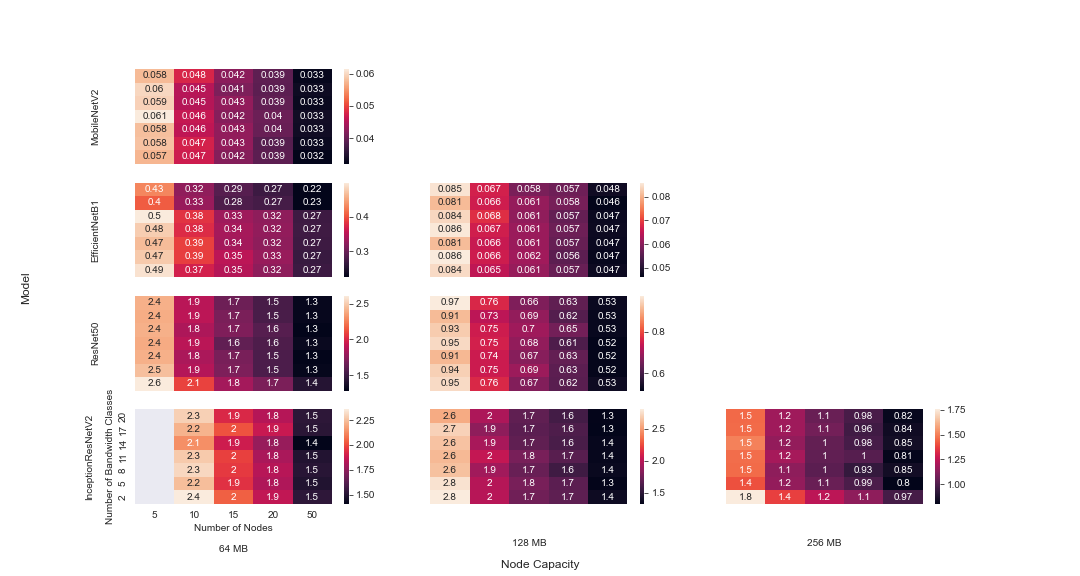}}
\caption{Color Map of Bottleneck Latency ($s$) based on Model, Node Capacity, Number of Nodes, and Number of Bandwidth Classes - Optimal Partitioning/Placement}
\label{fig:bottleneckplotsoptimal}
\end{figure}

In Figure \ref{fig:bottleneckplotsoptimal}, the lack of bottleneck latency values for InceptionResNetV2 with 5 nodes and 64MB node capacity indicates that the model could not be partitioned with these physical constraints.

In Figure \ref{fig:bottleneckplotsoptimal}, the color map was only generated for the node capacities which were too small for the models to fit on a single device of that capacity. All models were able to fit on a single 512 MB device.

For each model, the lowest bottleneck latency for a given node capacity comes from the combination of the most number of bandwidth classes and number of nodes. The lowest bottleneck latency comes with the highest node capacity. These results follow from the fact that a larger node and number of nodes allows the partitioning algorithm to have greater choice in selecting the smallest transfer sizes. Similarly, a high number of bandwidth classes allows the placement algorithm to better perform the $k$-path matching.
 
\begin{figure}[htbp]
\centerline{\includegraphics[scale=0.3]{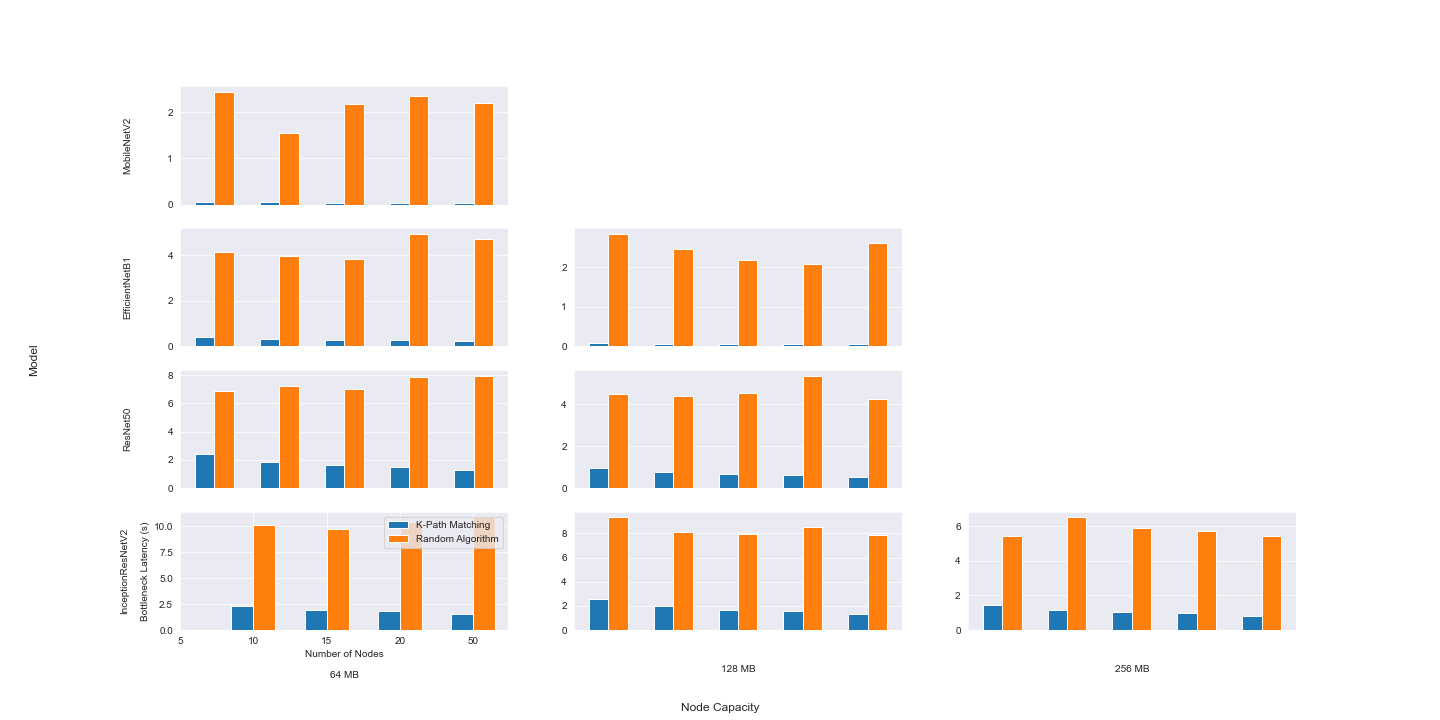}}
\caption{Comparison of Algorithm \ref{alg:kpathmatching} with Random Algorithm - based on Model, Node Capacity, Number of Nodes}
\label{fig:kpathvsrandom}
\end{figure}

In Figure \ref{fig:kpathvsrandom}, the optimal algorithm produces ~40x lower bottleneck latency than the random algorithm for MobileNetV2 on different node configurations. The difference is the smallest for ResNet50, with the optimal algorithm producing a ~2x lower bottleneck latency. For this selection of models, the optimal algorithm reduces bottleneck latency by ~10x on average. The models with the greatest variance in transfer size (see Section \ref{section:transfersizedist}) will result in the largest difference in bottleneck latency between the optimal random algorithms. Overall, we see that the optimal algorithm produces a significant reduction in bottleneck latency compared to the random algorithm.

\begin{figure}[htbp]
\centerline{\includegraphics[scale=0.3]{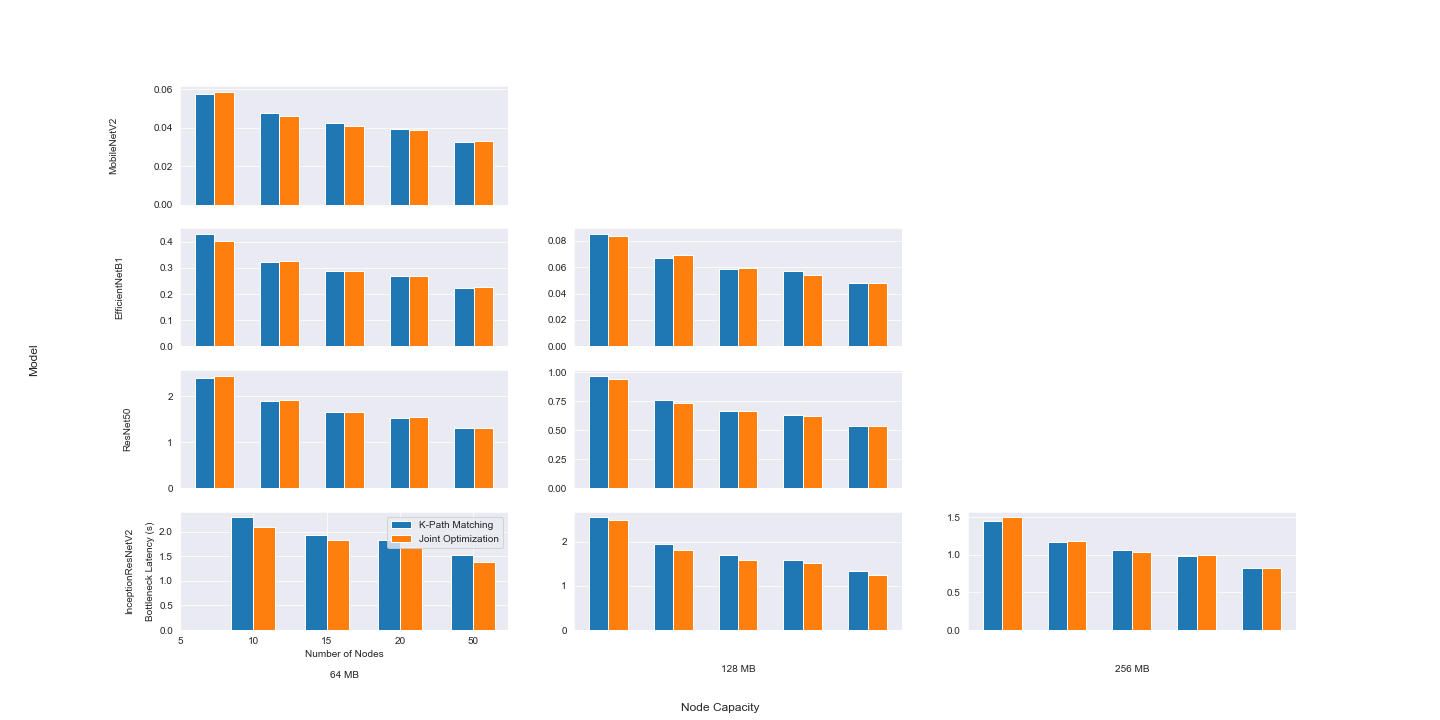}}
\caption{Comparison of Algorithm \ref{alg:kpathmatching} with Joint Optimization - based on Model, Node Capacity, Number of Nodes}
\label{fig:kpathvsjoint}
\end{figure}

\begin{table*}[!htp]\centering
\caption{Comparison of Approximation Ratios for K-Path Matching vs. Joint Optimization on Keras Pretrained Models}\label{tab:kpathvsjoint_edgecase}
\begin{tabular}{|c|c|c|c|}
\hline
&K-Path Matching &Joint-Optimization\\
\hline
16 MB &1.45 &\textbf{1.12} \\
\hline
32 MB &1.19 &\textbf{1.07} \\
\hline
64 MB &1.09 &\textbf{1.08}\\
\hline
\end{tabular}
\end{table*}

In Figure \ref{fig:kpathvsjoint}, the joint optimization algorithm tends to perform better for a smaller number of nodes. Since each of these algorithms use the same optimal partitioning logic, we can only compare the models based on their differing placement logic. As the number of nodes increases, our $k$-path algorithm performs better. This makes sense, because the difference in the greedy strategy of the joint optimization algorithm and the matching strategy of our algorithm only becomes more apparent as the communication graph grows bigger and there are more options for node paths. In particular, for 50 nodes, our algorithm outperforms the joint optimization algorithm by 35\%. We hypothesize that this trend would continue for more complex models which have a greater number of candidate partition points and a greater variance in transfer size, necessitating the $k$-path matching strategy to minimize bottleneck latency.

In Table \ref{tab:kpathvsjoint_edgecase}, the joint-optimization algorithm outperforms the k-path matching algorithm, although the different gets closer as the node capacity increases.

\subsection{Test Environment Results}

\begin{table*}[!htp]\centering
\caption{Fault Tolerance of Different Systems}\label{tab:failuretolerance}
\begin{tabular}{|c|c|c|c|}
\hline
&Our framework &Couper &DEFER\\
\hline
System IO fault-tolerance &$\times$ &$\times$ &$\times$\\
\hline
Network fault-tolerance &$\times$ &$\times$ &$\times$\\
\hline
Single node fault-tolerance &$\times$ &$\times$ & \\
\hline
Multi-node fault-tolerance &$\times$ & & \\
\hline
\end{tabular}
\end{table*}

\begin{table*}[!htp]\centering
\caption{Throughput and End-to-End Latency based on Graph Shape}\label{tab:testenvresults}
\begin{tabular}{|c|c|c|c|}
\hline
Number of Nodes &Graph Shape &Inference Throughput (Hz) &End-to-End Latency (s)\\
\hline
5 &Ring &0.072 &23.55 \\
\hline
5 &1x5 Grid &0.113 &15.86 \\
\hline
\end{tabular}
\end{table*}

Table \ref{tab:failuretolerance} compares our framework to the closely-related work Couper and our prior work DEFER. We see that since DEFER is simply a multi-threaded Python runtime, it has network and system IO fault-tolerance. However, since Couper is also a container orchestration framework, it additionally has single node fault-tolerance but cannot scale to multi-node fault tolerance because it is designed to be deployed on clusters with few edge devices. However, our framework is designed to be run on large edge clusters and has been tested with clusters of up to 50 nodes. Our framework only requires a collective cluster memory equivalent to the model partitions' memory and a single node's worth of storage space for the NFS server which houses the model partitions.
In Table \ref{tab:testenvresults}, we find the inference throughput and end-to-end latency for different cluster sizes and graph shapes. We see that a grid shape, due the closeness of its nodes, outperforms the ring shape for both inference throughput and end-to-end latency. \textbf{This table will be updated in a future version to include results for 9 node and 20 node configurations.}

\section{Conclusion}
We have presented a framework to partition and place a model across a set of resource-constrained edge devices, with the goal of maximizing inference throughput. We leverage containerization to increase robustness, scalability, and fault-tolerance of the system. 

We show that given certain characteristics about edge devices on a WiFi network, we can infer details about the communication graph and hardware requirements of most image and text models.

We find that we can reduce the bottleneck latency by 10x over a random algorithm and 35\% over a greedy joint partitioning-placement algorithm, although the joint-partitioning algorithm outperforms our algorithm in most practical use-cases. Furthermore we find empirically that for the set of representative models we tested, the algorithm produces results within 9.2\% of the optimal bottleneck latency. In our tests on our virtual cluster environment, we observed that our system has multi-node fault-tolerance as well as network and system IO fault-tolerance.

Our code is publicly available to the research community at \url{https://github.com/ANRGUSC/SEIFER}.

\subsection{Future Work}
With minor edits, we could extend our framework to work with geographically-distributed edge devices for a truly scalable edge inference solution.

Our results from Section \ref{section:partitionmem} suggest that with software changes, we could potentially run the average image model on a cluster of micro-controllers. We could use RiotOS \cite{riotos} without any containerization and perform optimizations to run with limited device memory. Some devices we could potentially take advantage of are the Raspberry Pi Pico \cite{rpi_pico} and Arduino Uno \cite{arduinouno}.

Secondly, as model size grows, the model partitions and their weights may occupy more storage space than a node's capacity. In the future, we could explore a more complex NFS database which uses shards across multiple nodes.

Finally, our results from Section \ref{section:partitionmem} suggest that we could run parallel streams of our inference framework on the same cluster. To do this, we would use different pod namespaces and cluster RBAC (role-based access control) for different instances of the system.

\section{Acknowledgements}
We would like to acknowledge the helpful input and pointers provided by Prof. Anil Vullikanti from the University of Virginia, particularly in directing us to the color-coding $k$-path algorithm. 

\bibliography{main}
\bibliographystyle{acm}
\end{document}